\documentclass[5p,times]{elsarticle}

\usepackage{amsmath}
\usepackage{amsthm}
\usepackage{algorithmic}
\usepackage{array} 
\usepackage{fixltx2e}
\usepackage{url}
\usepackage{booktabs}
\usepackage{moreverb,url}
\usepackage[colorlinks=false,bookmarksopen,bookmarksnumbered,hidelinks]{hyperref}
\usepackage{caption}
\usepackage{subcaption}
\usepackage{tabularx}
\usepackage{float}
\usepackage[table,xcdraw]{xcolor}
\usepackage{enumitem}
\usepackage[T1]{fontenc}
\usepackage{colortbl}
\usepackage{orcidlink}
\usepackage{cleveref}
\usepackage{makecell}
\usepackage{longtable}
\usepackage{multicol}
\usepackage{flushend}
\usepackage{threeparttable}

\graphicspath{{images/}}

\crefformat{section}{\S#2#1#3}
\crefformat{subsection}{\S#2#1#3}
\crefformat{subsubsection}{\S#2#1#3}

\newcommand{\SSr}[1]{\texttt{#1}}

\newtheorem*{innerthm}{\innerthmname}
\newcommand{\innerthmname}{}% initialize
\newenvironment{hypothesis}[1]
 {\renewcommand{\innerthmname}{#1}\innerthm}
 {\endinnerthm}
 
\newcolumntype{C}[1]{>{\hsize=#1\hsize\centering\arraybackslash}X}%
\newcolumntype{L}[1]{>{\hsize=#1\hsize\raggedright\arraybackslash}X}%

\begin{document}

\begin{frontmatter}

\title{A Generalized Feature Model for Digital Twins}

\author[docs]{Philipp Zech}
\ead{philipp.zech@uibk.ac.at}
\author[docs]{Yanis Mair}
\ead{yanis.mair@uibk.ac.at}
\author[docs]{Michael Vierhauser}
\ead{michael.vierhauser@uibk.ac.at}
\author[iese]{Pablo Oliveira Antonino}
\ead{pablo.antonino@iese.fraunhofer.de}
\author[iese]{Frank Schnicke}
\ead{frank.schnicke@iese.fraunhofer.de}
\author[aston]{Tony Clark}
\ead{tony.clark@aston.ac.uk}
\affiliation[docs]{
    organization={Department of Computer Science, University of Innsbruck},
    addressline={Technikerstrasse 21a}, 
    city={Innsbruck},
    postcode={6020}, 
    state={Tyrol},
    country={Austria}
}
\affiliation[iese]{
    organization={Department of Virtual Engineering, Fraunhofer IESE},
    addressline={Fraunhofer-Platz 1}, 
    city={Kaiserslautern},
    postcode={67663}, 
    state={Rhineland-Palatinate},
    country={Germany}
}
\affiliation[aston]{
    organization={School of Engineering and Applied Science, Aston University},
    addressline={1 Lister Str.}, 
    city={Birmingham},
    postcode={B4 7ET}, 
    state={West Midlands},
    country={United Kingdom}
}

\begin{abstract}
The adoption of Digital Twin technologies is rapidly expanding in diverse industrial, economic, and 
societal domains. Over the past decade, a multitude of studies, surveys, and investigations have 
been conducted, examining the nature, applications, and advantages of Digital Twins. However, up until 
now, no proposal for a comprehensive feature model exists that effectively captures the \emph{mandatory} 
and \emph{optional} features of Digital Twins. To address this shortcoming, in this article, we present 
a \emph{general feature model} for Digital Twins. Based on a systematic mapping study of existing literature, 
we developed a generalized feature model for Digital Models, Shadows, and Twins. To assess the validity of our 
proposed feature model, we have applied them to three use cases from the emergency, vehicular, and manufacturing 
domain. We conjecture that our proposed general feature model advances the field around Digital Twins by 
facilitating informed decision-making during design, enabling improved model-driven development of Digital 
Twins, and, eventually, fostering verification~\&~validation of Digital Twins by delivering a model-based 
foundation for test case inference.
\end{abstract}

%%Research highlights
\begin{highlights}
\item A generalized feature model (GFM) for Digital Twins (DTs) is proposed, derived from a systematic 
    literature review across six application domains.
  \item The model distinguishes Digital Models, Shadows, and Twins, offering semantic clarity and structured 
    classification of mandatory and optional features.
  \item The GFM aligns with Wagg’s DT maturity model, supporting progressive capabilities from monitoring to 
    autonomous control.
  \item Validated through three domain-specific use cases, the model demonstrates applicability and utility in 
    emergency services, vehicular systems, and manufacturing.
  \item The methodology follows the Design Science Research (DSR) paradigm and supports model-driven 
    engineering (MDE) workflows via formalized feature mappings.
\end{highlights}


\end{frontmatter}

\section{Introduction}
\label{sec:introduction}

\begin{figure}[htb]
    \centering
    \includegraphics[width=.45\textwidth]{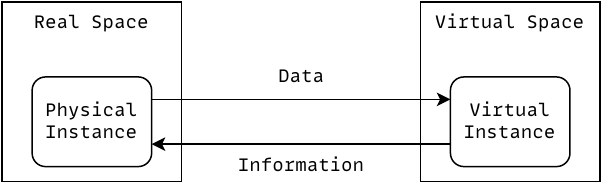}
    \caption{DT conceptual model proposed by Grieves and Vickers~\cite{kahlen_digital_2017}.}
    \label{fig:dt-conceptual}
\end{figure}

Digital Twins (DT) were originally conceived by Michael Grieves during a 2002 presentation 
at the University of Michigan to refer to a virtual representation of a physical system or process\footnote{https://www.challenge.org/insights/digital-twin-history}. 
In 2016 Grieves~\&~Vickers~\cite{kahlen_digital_2017} eventually set forth a seminal definition of DTs by 
describing them as digital information constructs about a physical object, system, or process that comprises
\begin{itemize}
        \item a \emph{virtual instance}, i.e., the assemblage of structure, 
                simulation, optimization, and operating models which together 
                replicate the physical instance, and
        \item \emph{interchanged data and connections} for bidirectional 
                communication between the virtual and physical instance,                 
\end{itemize} 
and usually are linked with the physical instance (cf.~Fig.~\ref{fig:dt-conceptual})
throughout its entire life cycle~\cite{kahlen_digital_2017}. 

\begin{figure}[htb]
        \centering
        \includegraphics[width=\columnwidth]{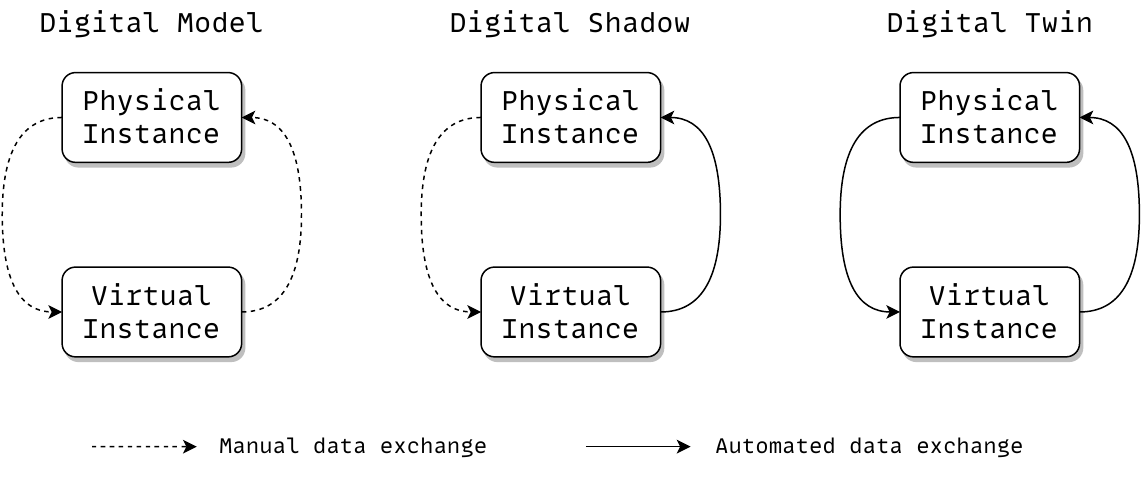}
        \caption{Types of DTs given the automation of data interchange as indicated 
        by dashed (manual) and automated (solid) arrows~\cite{kritzinger_digital_2018}.}
        \label{fig:types-of-twins}
\end{figure}

Typically, DTs are created by capturing real-time data from sensors, devices, and other sources and then 
simulating and mirroring the behavior, characteristics, and conditions of the physical entity in a digital 
environment. However, DTs do not merely represent  a collection of 3D models or simulations, but they incorporate 
real-time data, machine learning, and other technologies to 
provide a near-real-time representation of the physical entity~\cite{uhlemann2017digital,rasheed2020digital}. 
This implies that DTs employ various levels of \emph{twinning}, i.e.,~the degree of data interchange between the 
virtual and physical instance, which led to Kritzinger et al.' extension of the concept of a DT towards two 
related concepts, viz.~\emph{Digital Model} (DM) and \emph{Digital Shadow} (DS) as depicted in Fig.~\ref{fig:types-of-twins} 
to more accurately capture the landscape of different types of DTs w.r.t.~their purpose~\cite{kritzinger_digital_2018}.

\begin{figure}
\centering
    \includegraphics[width=\columnwidth]{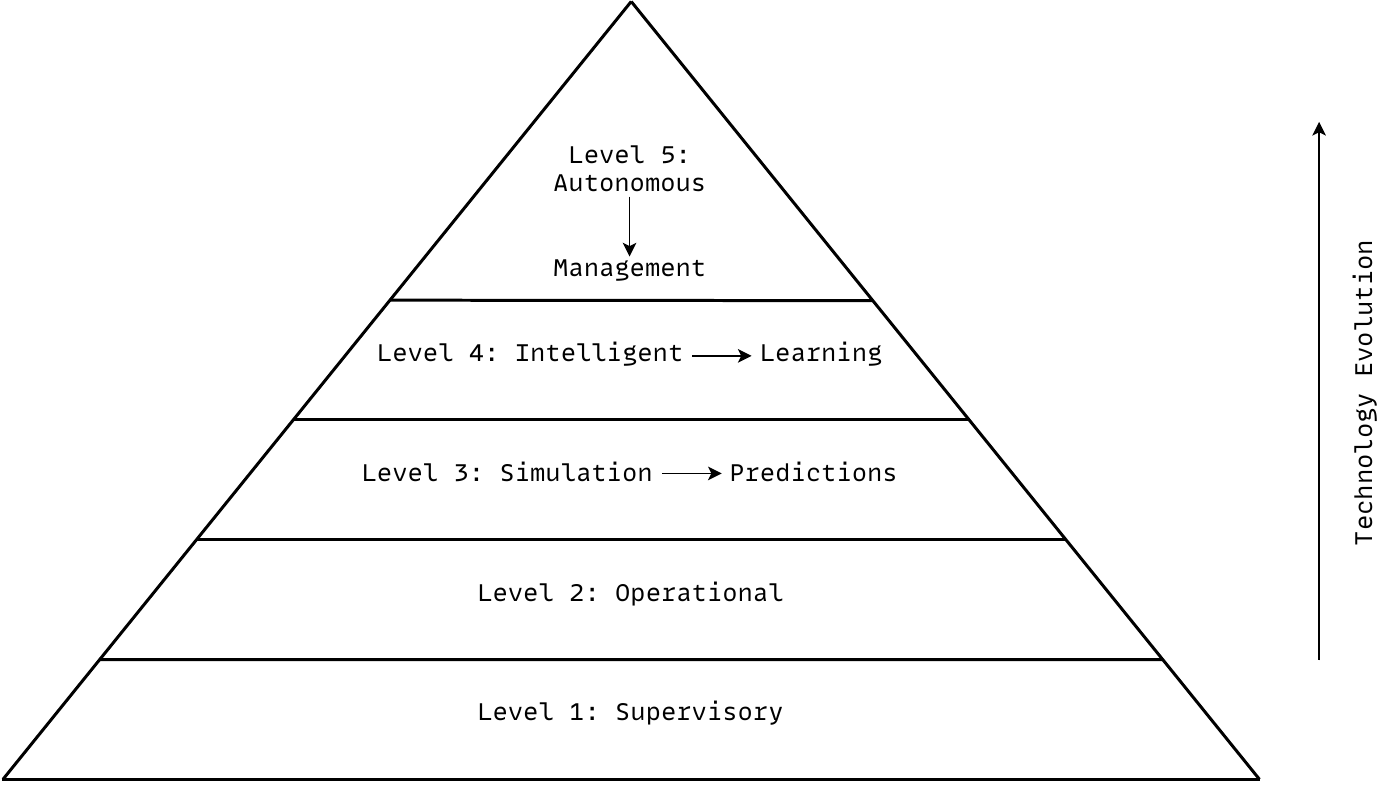}
    \caption{DT maturity levels~\cite{wagg2020digital}.}
    \label{fig:maturity}
\end{figure}

When considering the term DT, it is important to recognize that there are both different types of \emph{application domains} 
and \emph{twin functionality}. Wagg et al.~\cite{wagg2020digital} have argued that there is a hierarchy of \emph{capabilities} 
providing a useful lens through which DTs  can be assessed w.r.t.~their maturity (cf.~Fig.~\ref{fig:maturity}):
\begin{description}
    \item[Level 1] simply measures variables from the real world, it may detect when the 
        measures reach certain values or ranges, but does not link them together or interpret 
        collections of values. 
    \item[Level 2] is model-based and goes beyond simple measurement collection, interpreting 
        collections of values as holistic situations. At this level, DTs are often implemented 
        as dashboards for human-centric situation assessment. 
    \item[Level 3] enables the current situation to be projected into the future using simulation 
        principles as a component of \emph{what-if} analysis which may be accomplished through 
        a degree of model configuration. Typically, the execution rules are stochastic, 
        as various progression options may be available for each system state. Level 3 can, 
        therefore, provide more sophisticated decision support by presenting options and alternatives 
        to human decision-makers and utilizing simulation techniques to recommend the best 
        option described by desired outcomes. 
    \item[Level 4] utilizes machine learning and AI to automate Level 3 decisions normally 
        made by humans. Execution models can be pre-trained to make the correct decisions, 
        or they can dynamically adapt to improve and become more likely to make correct 
        decisions. Level 4 learning may still necessitate some level of human intervention, 
        but it can reduce the burden of oversight required to ensure that the actual system 
        is behaving as desired. 
    \item[Level 5] removes the burden of decision-making from humans by automating the control 
        of the real-world system by the DT. To achieve this, the level 5 DT must maintain an 
        accurate model of the system that has been learned through training or adaptation, and 
        make decisions through forward exploration of potential future states measured against 
        system goals.
\end{description}
From a feature perspective, these maturity levels provide a first glimpse into 
relevant features a DT must comprise to implement a necessary capability.  These levels 
eventually define necessary requirements to fulfill for implementing a related capability. 
This idea of taking a feature-oriented stance is highly reminiscent of feature-driven development~\cite{palmer2001practical} 
(FDD), an agile software development methodology for long-term, complex projects that capitalizes 
on the initial establishment of a feature model from given requirements. 
Subsequently, this feature model defines the roadmap for implementation to eventually arrive 
at the anticipated product. Frankly, we are also not the first to speculate about the application 
of FDD for the purpose of engineering a DT~\cite{ren2022feature}. However, what has not been 
investigated so far is the establishment of a \emph{generalized feature model} for DTs that allows 
feature description at an abstract level, specifically pertaining to \emph{mandatory} and \emph{optional} 
features for informed decision-making during design, development, and verification~\&~validation (V\&V) of a DT. 
Typically, before constructing a DT, two important early decisive actions must be taken, viz. 
(i) eliciting the purpose of the DT, e.g., maintenance, optimization, or failure detection, and (ii) defining 
the goal to be achieved, e.g., doubling performance or preventing errors. 

Given this, stakeholders then need to identify the physical entity they need to digitally replicate and 
subsequently collect data from relevant sources, including sensors, devices, and other systems. This 
information is then organized in a DT platform. To provide a more complete and accurate representation, 
the platform may also incorporate data from other sources, such as weather forecasts and consumer behavior 
patterns~\cite{chambers2024self}. We conjecture that the provisioning of a \emph{generalized feature model} 
for feature-driven development of DTs that captures these two fundamental design decisions would greatly 
benefit the research community and stakeholders in fully understanding the concept (of DTs, respectively), 
how to approach their design, deliver given requirements by implementing relevant features and, finally, 
verify~\&~validate the DT in its environment against the initially defined requirements.

Aside from the above discussion, we further argue that there are pressing business drivers for a 
structured and systematic approach to engineering DTs. For instance, as of 2022, the market share of DTs 
has only reached \$ 7 billion dollars, with an expected share of \$ 183 billion dollars in 2031~\cite{gartner2022}. 
Reaching this market share, however, can only be achieved by 
introducing novel structured and systematic approaches when implementing a DT to achieve the anticipated 
return on investment~\cite{zech2025empirical}.

\subsection{Challenges and Contributions}

Commensurate with the above, in our work, we introduce a \emph{generalized feature model} (GFM) for 
DTs for enabling informed decision-making during core stages of DT creation: design, implementation, and 
validation~\&~verification. Specifically, we conduct a systematic literature review to infer a GFM 
for DTs from existing literature, thereby tackling the following challenges, viz., a
\begin{enumerate}
    \item lack of a comprehensive understanding of the DT problem space (cf.~\cref{ssec:problem-space}),
    \item lack of an understanding of the \emph{mandatory} and \emph{optional} features of DTs
        and, consequently, the current
    \item lack of a \emph{generalize feature model} for DTs that captures the problem space, 
        by \emph{mandatory} and \emph{optional} features of DTs,
\end{enumerate}
thereby delivering the following contributions, viz.,
\begin{enumerate}
    \item an exploration of the DT's problem, design, and solution space,
    \item a concerted terminology for DTs across domains and feature dimensions,
    \item an elicitation of \emph{mandatory} and \emph{optional} features of DTs, and
    \item systematically established \emph{generalized feature model} for DTs capitalizing on contributions 1-3.
\end{enumerate}
Specifically, we will investigate the following research questions as part of our contribution, viz.,
\begin{itemize}
  \setlength\itemsep{0.15em}
    \item[\textbf{RQ1}] What are the DT problem spaces?
    \item[\textbf{RQ2}] What are the DT design spaces?
    \item[\textbf{RQ3}] What are \emph{mandatory} features of DTs across domains?
    \item[\textbf{RQ4}] What are \emph{optional} features of DTs across domains?
    \item[\textbf{RQ5}] What is the DT solution space?
\end{itemize}

Our research employs the Design Science Research (DSR) methodology~\cite{wieringa_dsr_2014} to develop 
artifacts aimed at facilitating the construction of DTs. The centerpiece of our contribution is the GFM, 
developed through a systematic literature survey to infer features relevant to the GFM. These artifacts 
address the following design science problem, articulated using the DSR problem 
template~\cite{wieringa_dsr_2014}:
\begin{table}[H]
  \small
	\begin{tabularx}{\columnwidth}{@{}lX@{}}
	\toprule
	\textbf{Improve} \emph{DT engineering} (Context) \\
	\textbf{by eliciting} \emph{mandatory and optional features of DTs} (Artifact) \\
	\textbf{by mapping} \emph{the DT problem space to features} (Requirement) \\
	\textbf{to deliver} \emph{a generalized feature model for DTs}. (Goal) \\
	\bottomrule
	\end{tabularx}
\end{table}

Our work is based on the conjecture that given the problem space (cf.~\cref{ssec:problem-space}), 
relevant concepts comprising the DT design space can be identified (cf.~\cref{ssec:design-space}) 
and aligned with Wagg's maturity model (cf.~Fig.~\ref{fig:maturity}. This alignment 
exhibits that DTs conceptually are limited in what they can solve thus motivating the 
postulation of the following hypothesis
\begin{hypothesis}
{Hypothesis}[H1]
    There exists a GFM for DTs
\end{hypothesis}
which we investigate against the obvious null hypothesis
\begin{hypothesis}{Hypothesis}[H0]
    There exists no GFM for DTs.
\end{hypothesis}

Via a systematic literature review, we propose such an initial feature model (cf.~\cref{sec:methodology}) 
which we then evaluate on selected DT instances to assess the model's suitability and fitness. 

\subsection{Scope and Limitations}

Our study will adopt a rather broad and flexible approach. Due to the "neophytic and pervasive" 
nature of DTs in that they are on the brink of "invading" multiple domains, cf.~construction 
engineering~\cite{jiang2021digital}, manufacturing~\cite{soori2023digital}, health-care~\cite{sun2023digital}, 
socio-economics~\cite{yossef2023social}, power grid optimization~\cite{jafari2023review} and many more, 
our proposed GFM needs the necessary level of \emph{generalization} to be applicable across all these 
domains. In addition, we will follow a slightly adapted working definition of DTs compared to the 
initially cited one of Grieves~\&~Vickers, Following Catapult who defines a DT as \emph{a live digital 
coupling of the state of a physical asset or process to a virtual representation} (comprised of 
models, respectively) \emph{with a functional output}\cite{catapult2021untangling}. 
Compared to Grieves~\&~Vickers definition, Catapult additionally emphasizes the aspect of 
functionality which, from our point, is crucial in that a DT follows a dedicated purpose. 
Finally, as for the life cycle of a DT, our proposed GFM will cover the design, development, and 
validation~\&~verification phases of a DT. During design, the GFM facilitates informed 
decision-making. During development, if following a model-driven approach (cf.~Trujillo et al.~\cite{trujillo2007feature}, 
the GFM facilitates the inference of modeling artifacts using model-2-model or model-2-text transformations~\cite{kahani2019survey}. 
Finally, during the validation~\&~verification phase, the GFM facilitates test case development 
by providing a structured foundation for test case design from features.

\vspace{.25cm}
\textbf{Organization:} \cref{sec:background} discusses related work and establishes the context 
of our study by addressing RQ1 and RQ2. \cref{sec:methodology} outlines our research methodology, 
followed by \cref{sec:models} which presents our GFM for DTs, adressing RQ3-RQ5.  
\cref{sec:application} then exemplifies the application of our GFM for various tasks, for example,  
inferring a development road map, test scenarios, classifying and comparing DTs, and more. 
\cref{sec:discussion} discusses our results, viz.~the empirically assessed feature model 
for DTs and its relevance for advancing the field. We conclude in \cref{sec:conclusion} 
with a summary of the main implications of our study and future work.
\section{Background and Related Work}
\label{sec:background}

In this section, aside from discussing related work on DT feature models (cf.~\cref{ssec:rel-work}) 
we address the first two research questions, RQ1 and RQ2, by discussing the DT problem space (RQ1 -- cf.~\cref{ssec:problem-space}) 
and design space (RQ2 -- cf.~\cref{ssec:design-space}). The necessity for such 
in-depth discussion of problem and design spaces is vital in fully understanding the concept of a DT 
and its potential application.

\subsection{Digital Twin Problem Space: Challenges and Considerations}
\label{ssec:problem-space}

\begin{figure}[htb]
    \includegraphics[width=\columnwidth]{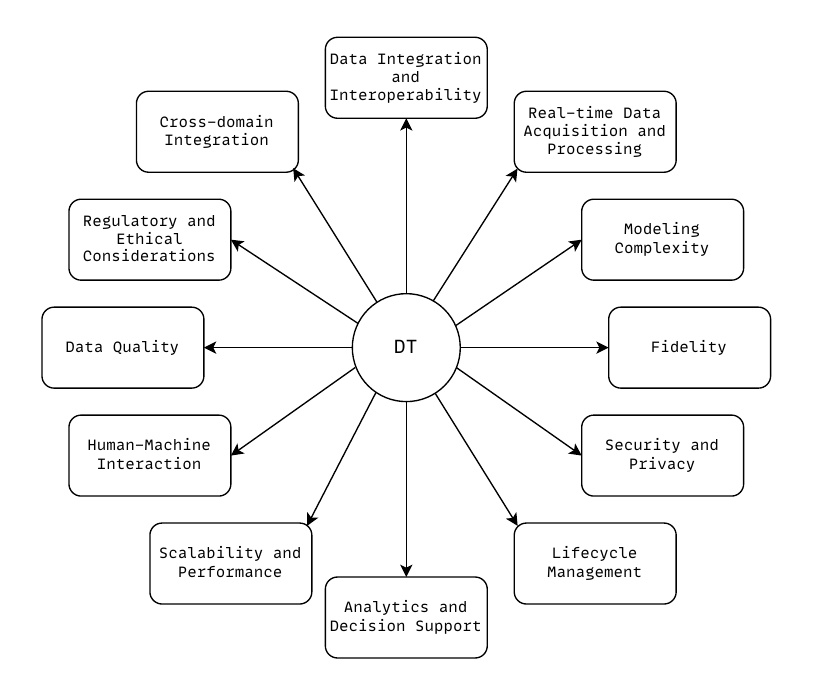}
    \caption{Challenges and considerations rooted in DT problem spaces.}
    \label{fig:problem-space}
\end{figure}

Traditionally, a problem is defined as a situation where a clear goal (e.g., building and running a DT) 
exists, but there is no obvious or easy way of attaining said goal. A key idea is to view problem-solving 
as a search through a "problem space" with the purpose of finding a way to the goal. The problem space then 
is denoted by the set of all possible states, operators, and paths that can be considered or explored in the 
process of solving a particular problem, viz.~\cite{10.7551/mitpress/3966.003.0013}:
\begin{itemize}
  \setlength\itemsep{0.15em}
    \item \textbf{Initial State}: The starting point of the problem-solving process.
    \item \textbf{Goal State}: The desired outcome of the problem-solving process.
    \item \textbf{State}: The conditions or configurations that the problem can be 
        in at any point during the problem-solving process.
    \item \textbf{Operators (or actions)}: The steps that can be taken to move from 
        one state to another within the problem space and are defined based on the 
        problem domain and the rules governing the system (cf., development and 
        integration of features in the context of a DT).
        in the context.
    \item \textbf{Paths}: A sequence of operators that lead from the initial state 
        to the goal state. Finding an optimal or feasible path is often the primary 
        objective of problem-solving. If not feasible, another solution is to \emph{satisfice}~\cite{10.7551/mitpress/3966.003.0013}: 
        do not strive to achieve the best solution, but instead a satisfactory one.
\end{itemize}
To develop a feature model for DTs, a thorough assessment of the 
challenges and considerations along a solution path in the problem space is necessary 
as outlined in Fig.~\ref{fig:problem-space} and discussed in the following.
Observe that in the case of DTs, we explicitly consider \emph{problem spaces} as it is neither feasible nor 
rational to define one \emph{single} problem space that captures all 
of the possible use cases, applications, and resulting manifestations of DTs.

\subsubsection{Data Integration and Interoperability} An essential obstacle is the integration of data from 
diverse sources (sensors, IoT devices, operational systems) into a cohesive data model for the DT. This 
entails resolving concerns related to accuracy, consistency, frequency, modality, and uniformity of data 
to guarantee smooth compatibility, integration, and provisioning among different systems.

\subsubsection{Real-time Data Acquisition and Processing} Ensuring a precise depiction of the state of the 
physical system necessitates ongoing data updates. Managing substantial amounts of runtime data and effectively 
processing it to promptly update the DT poses notable technical obstacles.

\subsubsection{Fidelity}
Determining the appropriate level of fidelity required to meet the needs of the intended use case. Identifying 
which parts of the intended use case require which level of fidelity, resulting in a mixture of high-fidelity 
and low-fidelity aspects in a DT.

\subsubsection{Data Quality}
Ensuring that the data is of appropriate quality for the intended use case. Providing semantic descriptions and 
meta-data, for instance, measuring ranges and measuring resolutions.

\subsubsection{Modeling Complexity} Constructing a proficient virtual instance further requires developing 
intricate models that precisely reflect the activities and interconnections of the physical instance. Dealing 
with the intricacy of these models, which involve aspects like heterogeneity, multi-level structures and 
hierarchies, multi-scale dynamics, nonlinear behaviors, and uncertainties, presents a significant challenge.

\subsubsection{Security and Privacy} Cybersecurity concerns pose a significant risk to DTs by virtue of DTs' 
heavy reliance on data. It is essential to guarantee the protection, confidentiality, and accuracy of data 
at every stage of the DT's life cycle, especially when handling sensitive information from operational systems.

\subsubsection{Life cycle Management} As time progresses, DTs inevitably evolve and are subject to changes, 
thus requiring  updates to accurately represent modifications in the physical instance (cf.~\emph{as-is} 
modeling), such as enhancements, upkeep, or deterioration. Effectively managing the whole lifespan of DTs, 
which includes activities like version control, change management, and preserving historical data, is 
crucial for ensuring their usefulness in terms of reliability.

\subsubsection{Analytics and Decision Support} To derive relevant insights from DTs, it is necessary to 
employ sophisticated analytics approaches. The task of creating algorithms for predictive maintenance, 
optimization, anomaly detection, and decision support using data in the DT is highly challenging.

\subsubsection{Scalability and Performance} Ensuring scalability and performance of the underlying infrastructure 
becomes crucial as DTs are used in more intricate systems and employ growing datasets. This includes factors 
such as processing techniques and resources, network bandwidth, and latency.

\subsubsection{Human-Machine Interaction} Maximizing the usefulness and acceptance of DTs requires enabling 
effective human interaction through intuitive visualization, user interfaces, and analysis and collaboration 
tools. User interfaces should be designed in an inclusive manner to meet the needs and requirements of all 
participating stakeholders.

\subsubsection{Regulatory and Ethical Considerations} The utilization of DTs can raise both ethical and 
regulatory concerns, particularly when handling sensitive and confidential data or dealing with critical 
infrastructure. It is crucial to guarantee adherence to pertinent data governance standards and ethical 
principles.

\subsubsection{Cross-domain Integration} DTs frequently encompass various domains, such as manufacturing, 
healthcare, or smart cities, necessitating the integration of numerous disciplines and sectors. To fully 
harness the capabilities of DTs, it is crucial to overcome obstacles associated with domain-specific standards, 
terminology, and practices.

\subsection{Digital Twin Design Space: Concepts}
\label{ssec:design-space}

Given our discussions from the previous section, we now focus on investigating our second research question (RQ2), which 
elicits the DT design spaces. Our goal is to provide a thorough overview of the relevant concepts that 
are part of a DT's design spaces following our discussion of the DT problem spaces.

\begin{figure*}[htb]
    \includegraphics[width=\textwidth]{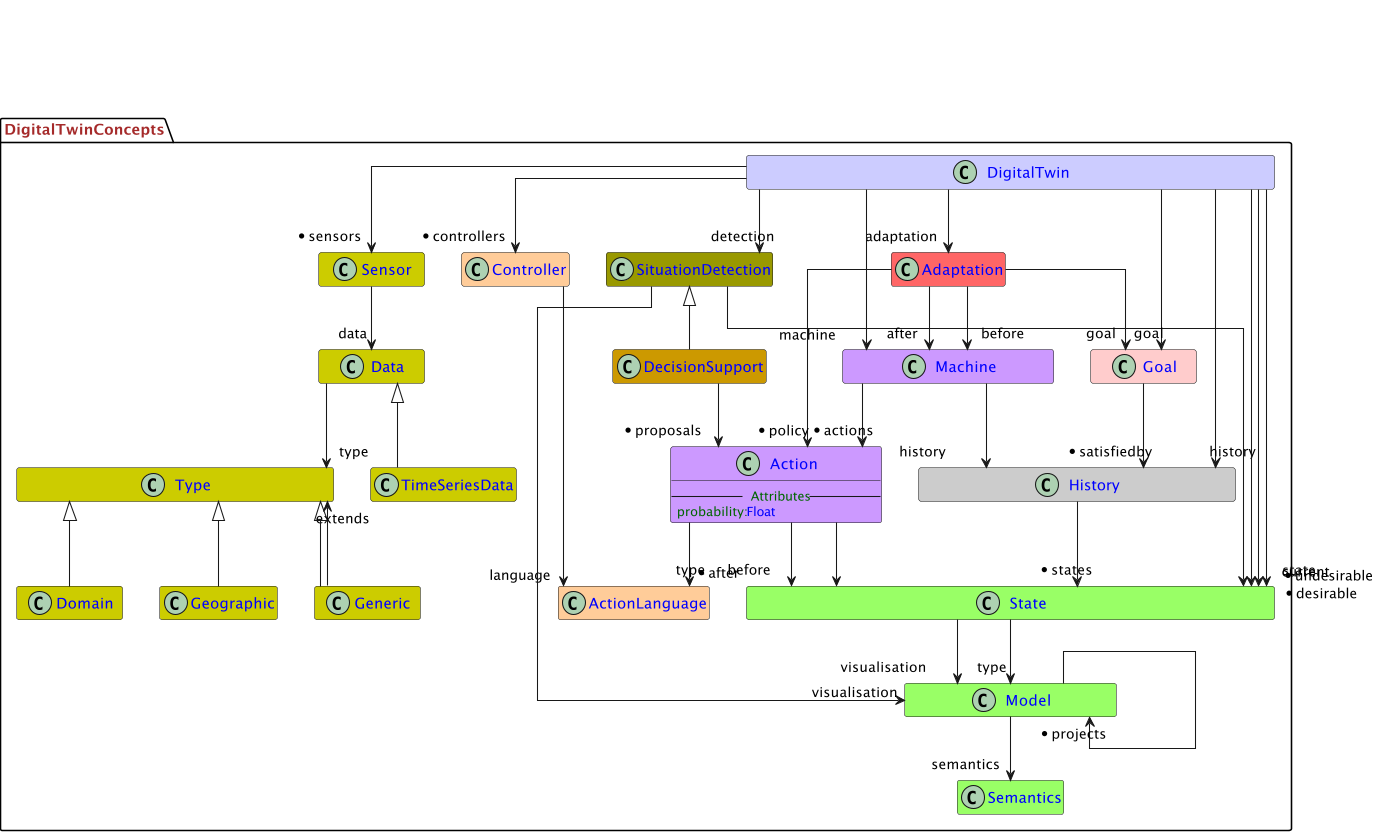}
    \caption{Concepts of the DT design spaces.}
    \label{fig:concepts}
\end{figure*}

Fig.~\ref{fig:concepts} shows a model of a set of concepts representing the design space for DTs. The 
concepts are grouped into {\it dimensions} some of which are {\it mandatory} and some of which are {\it optional} 
depending on the type of DT. This section provides an overview of the dimensions which are color-coded 
in Fig.~\ref{fig:concepts}.

\subsubsection{Monitoring}

A mandatory feature of a DT is that they do in fact monitor a real-world system or asset. This is 
reflected by the {\tt Sensor}  concept that provides {\tt Data} to the twin. The sensor data may be collected 
a-priory and then supplied to the twin, or supplied in real-time.

Sensor data is associated with a specific type. In an extensible twin or a digital twin platform, it is useful 
for the data to be in a general-purpose format (represented by \texttt{Generic}) and for the general format to 
be \emph{extensible} via more specific data type plugin-types. Domain-specific data types are often necessary, 
an example of such a type is shown as \emph{Geographic}.

It is not strictly necessary for a digital twin to have time series data, however, this is likely since sensors 
will collect real-world data over time and therefore can supply it as a stream of time-tagged data values to the 
twin.

\subsubsection{State}

A mandatory feature of a DT is that it maintains a representation 
of the real world. This is represented via the current \texttt{state} in the concept model. A state is an instance 
of some form of data \texttt{Model}.

The fidelity of the data model with respect to the real world is captured as a \texttt{projects} relationship 
on \texttt{Model}. The idea is that there is a model that is in one-to-one correspondence with the real-world 
data being measured by the sensors, however, the twin may be designed to represent only a subset of real-world 
information. In an extreme case, the twin may simply represent a collection of values (for example, the $(x,y)$ 
position of a vehicle). At the one end of the spectrum, a twin could represent a complete model of a vehicle 
including its position, direction, engine state, emissions, and the relationships between these features. 

A model will have a semantics that determines whether the current state constructed by the twin is {\it well-formed}. 
At the low-fidelity end of the spectrum, the semantics simply checks that variables have the correctly typed 
values. At the other extreme, the semantics will check the internal consistency of relationships across the 
state and may even check whether the current state is valid in terms of its history.

\subsubsection{Visualization}

The state managed by a twin is communicated to twin operators using a model 
(denoted \texttt{visualisation}). The concept involves a communication model that facilitates operator interaction 
with the twin, where the system's state in turn is mapped to this communication model. In many cases, a twin will 
be implemented with a user-facing dashboard, serving as a graphical representation of its model. This functionality 
provides the lowest level of a DT: one that manages data from the real world and displays it on a screen for human 
processing. Even within this level, there are variations depending on the fidelity of the state model and whether 
the visualization faithfully represents the semantics and time-labeled history.

\subsubsection{Intention}

A Digital Twin may include a \emph{Goal} which is a predicate over the history of states managed by the twin. 
The goal may be used in a variety of ways, for example, it could express desirable states for the twin to achieve 
or maintain. Alternatively, it could encode undesirable states which the twin aims to avoid.

\subsubsection{Situation Detection}

A twin may monitor the current state to detect situations (\texttt{SituationDetection}). When the situations 
are detected, they are flagged, and the twin operator is notified. This is achieved as a collection of predicates 
that apply to instances of the state model. The detection mechanism may have knowledge of the visualization model 
and thereby display the situations of interest graphically.

\subsubsection{Decision Support}

A twin may define actions that affect the internal state. Given such actions, an extension of situation-detection 
is decision-support which, given a situation and a desired goal, proposes an action that is most likely to achieve 
the goal (desired state or state-avoidance). Where multiple actions are possible, each is assigned a probability 
regarding its likelihood for success. Given a suitable visualization model, the optional actions can be presented 
to the user for their consideration.

\subsubsection{Behaviour}

A twin may be equipped with a {\tt Machine} consisting of a collection of state 
transformation actions. It is likely that the machine is stochastic in 
the sense that for any given state there are several possible actions of 
differing probabilities.

Twin behavior provides the basis for many different types of Digital Twin 
because it can be linked to sets of histories both in terms of how the twin 
got to the current state and what is possible from the current state. Note 
that not all executions that can be generated by a machine need to satisfy 
the goal of the Digital Twin.

\subsubsection{Simulation}

A machine is the basis of {\it what-if} analysis. Given a current state, 
is it possible to run the twin forward? Since the machine is stochastic, 
such runs will lead to many different histories; each history will be 
associated with a different probability.

Sets of histories can then be analyzed. It is possible to ask whether all 
the histories satisfy the twin goal, whether none of them do, the probability of, for example, 
selecting the right actions to achieve the goal or the shortest route to the goal.

Furthermore, linking the simulation to the visualization model allows the user of the 
DT to explore forward in time from the current state and to 
represent the resulting states graphically.

\subsubsection{Adaptation}

A machine represents a relationship between twin states. Adaptation is the 
process of transforming a relationship towards becoming a function (or 
equivalently reducing the amount of non-determinism in the relationship). 
Most machine-learning techniques can be viewed in these terms.

DT adaptation is applied to the behavior of the twin, in order 
to make it more likely that it will achieve its goal. The adaptation may 
be performed on historical sensor data or may be performed in real-time 
as data is received from the sensors, or can be performed as a mixture 
of both.

\subsubsection{Control}

Some DTs are attached to controllers. the controller sends 
actions in an \texttt{ActionLanguage} and applies the actions to the real world. This 
creates a measure-process-action loop between the real world and the DT.

Control may be integrated with any of the concepts outlined above. For 
example, a twin that provides decision support may allow a user to select 
an action that is then executed in the real world. For example, a twin may 
support simulation to select the most appropriate action to perform next. 
Adaptation may be used to incrementally reduce the list of available actions 
that are offered to the user. Finally, adaptation may be used to remove 
the user altogether so that the choice of the next action is fully automated.

\begin{figure}[htb]
\centering
    \includegraphics[width=\columnwidth]{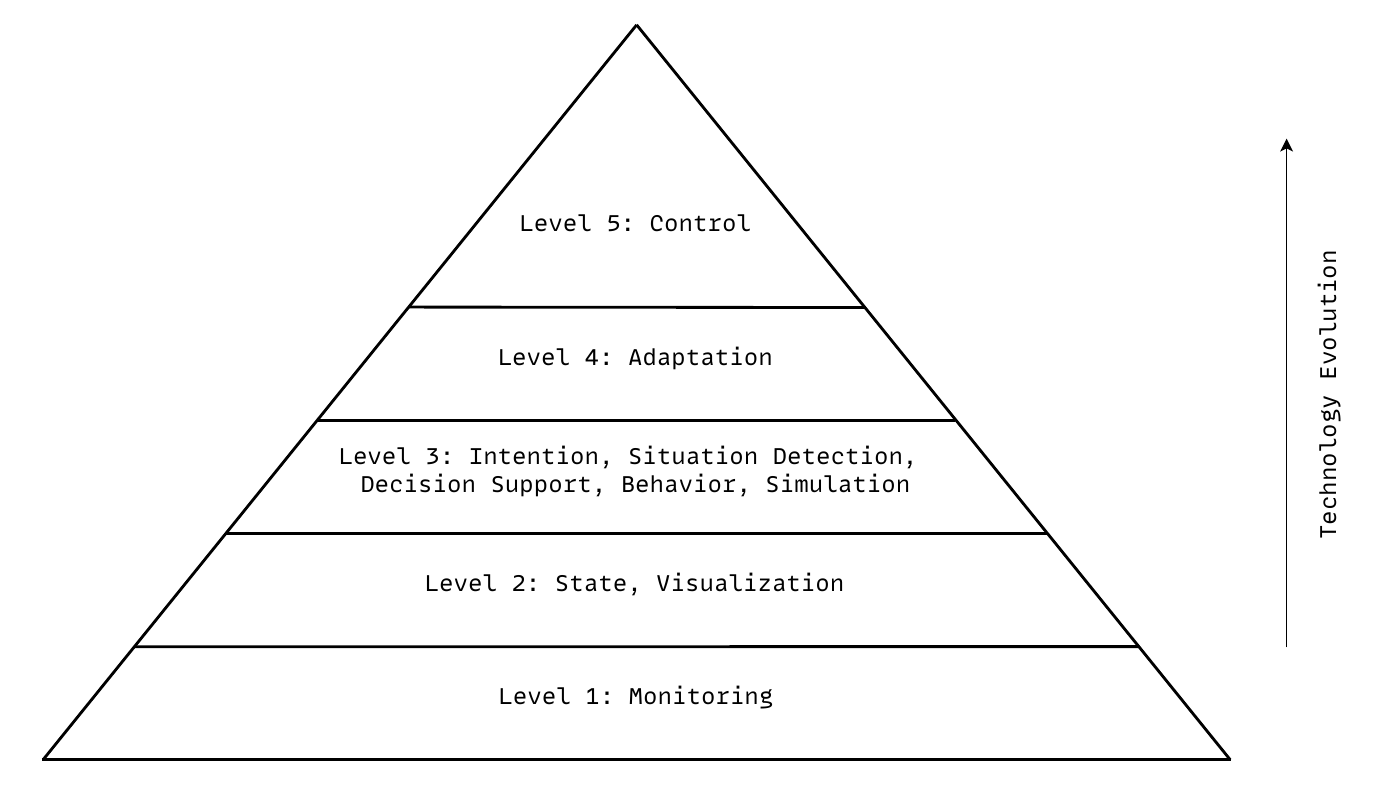}
    \caption{DT concepts aligned with Wagg et al.'~maturity levels~\cite{wagg2020digital} (cf.~Fig.~\ref{fig:maturity}).}
    \label{fig:concept_maturity}
\end{figure}

\vspace{.25cm}
Fig.~\ref{fig:concept_maturity} aligns these concepts with the earlier introduced maturity 
model of Wagg et al.~\cite{wagg2020digital}. This blatantly epitomizes the fitness and 
relevance of the proposed concepts. Further, it also demonstrates that a DT comprises more 
concepts than plain monitoring, and state representation and visualization. Crucially, the 
evolution from Level 2 to 3, according to Wag et al.~demarcates 
the advancement from a mere digital shadow to a DT (cf.~Fig.~\ref{fig:types-of-twins}, 
i.e., from plain monitoring and visualization towards higher-level capacities like simulation, 
improved decision-making, and autonomous control.

This alignment also fits very well with the Industry 4.0 maturity index \cite{becker2017industrie} 
suggested by Acatech where the stages of Industry 4.0 development are described. There, the following 
stages are described and correspond to the DT maturity levels as follows:
\begin{enumerate}
    \item Visibility, corresponding to Monitoring, State and Visualization
    \item Transparency, corresponding to Intention, Situation Detection and Decision Support
    \item Predictive Capacity, corresponding to Simulation
    \item Adaptability, corresponding to Adaptation and Control
\end{enumerate}
\subsection{Problem Space vs.~Design Space}

\begin{figure}[htb]
\centering
    \includegraphics[width=\columnwidth]{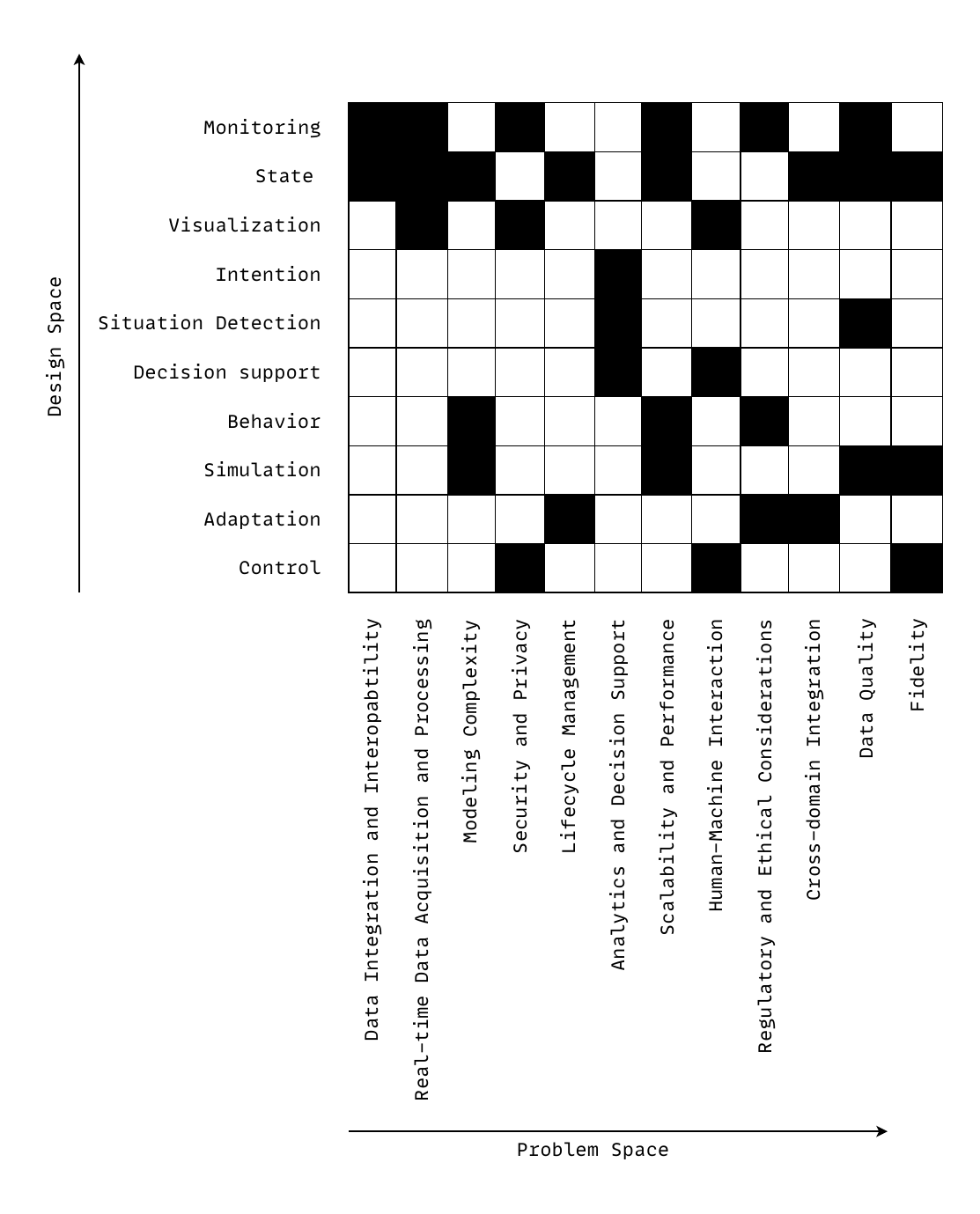}
    \caption{DT problem space (cf.~Fig.~\ref{fig:problem-space}) aligned with DT concepts (cf.~Fig.~\ref{fig:concepts}).}
    \label{fig:problem-design}
\end{figure}

Obviously, the problem and design space are directly connected, as the latter provides 
the landscape in which a solution can be found to a specific problem. 
Fig.~\ref{fig:problem-design} limns down this landscape as a 2D-map where elements from the 
design space (cf.~vertical dimension) are correlated with elements from the problem space 
(cf.~horizontal dimension). In that sense, Fig.~\ref{fig:problem-design} provides a high-level 
outline of which concepts might come in as a remedy to successfully navigate the 
problem space towards a feasible and satisfying solution.
Crucially, \cref{sec:models}, which introduces our GFM for DTs, refines the 
design space along concrete features, eventually delivering a set of "corrective actions" 
in terms of features to be implemented achieve an  envisaged solution.

\subsection{Related Work}
\label{ssec:rel-work}

Despite the rapid growth of DT-related research~\cite{yao2023systematic}, so far no attempts 
have been made to develop a feature model for DTs. However, as discussed in the following, efforts 
have been undertaken towards the establishment of analysis frameworks~\cite{autiosalo2019feature,boyes2022digital}, 
mapping of underlying software engineering trends~\cite{dalibor2022cross}, and the construction 
of maturity models~\cite{wagg2020digital,su13158224} to assess the capabilities of a DT in regard 
to its evolutionary stage (cf.~Kritzinger et al.~\cite{kritzinger_digital_2018}).

Autiosalo et al.~introduced a framework that utilizes features discovered through a literature analysis  
to structure industrial DTs in a feature-based manner~\cite{autiosalo2019feature}. As part of this, 
they describe 8 features, including data link, coupling, identification, security, data storage, user 
interface, simulation model, analysis, artificial intelligence, and computation. Their framework 
enables the universal definition and structuring of DTs based on three main principles. Firstly, 
all DTs are composed of a specific set of characteristics. Secondly, these characteristics can be 
used to compare different instances of DTs. Lastly, the characteristics can be combined through a 
data link feature to create future DTs more efficiently~\cite{autiosalo2019feature}.
Although we generally agree with these three propositions, we believe that their approach to feature 
definition for DTs is somewhat imprecise, requiring further refinement. The defined features remain 
vague and provide a rather high level of abstraction and 
conceptuality (cf.~\cref{ssec:design-space}). This level of abstraction fails to capture the 
distinctive \emph{user-visible aspects, qualities, or characteristics} that are considered as 
features~\cite{kyo1990foda}.

The analysis framework presented by Boyes and Watson seeks to provide a collection of characteristics 
for characterizing a DT, or comparing DTs with each other~\cite{boyes2022digital}. Through a thorough 
analysis of relevant literature, particularly focusing on architectural models and design approaches, 
Boyes and Watson leverage the proposed DT components by Catapult~\cite{catapult2021untangling}, viz.~live, 
digital coupling, state, physical asset or process, and virtual representation to propose four functional 
categories: digital coupling, tools, digital representation, and functional output. 
Furthermore, they provide a comprehensive examination of the functional aspects of each category, 
thus deducing a collection of functional characteristics for comparing DT capabilities. Similar to the 
proposal by Autiosalo et al., we argue that the range of high-level functional categories and their 
associated characteristics are still limited. Their work only examines the typical traits of DTs, cf.~their 
characteristics, without considering their unique physical attributes, cf.~the features\footnote{Longman 
Dictionary of Contemporary English}.

Dalibor et al.~performed a thorough and interdisciplinary systematic mapping study on software 
engineering for DTs~\cite{dalibor2022cross}. The article cross-examines six fundamental questions, viz.,~\emph{Who uses 
DTs for which purposes?}, \emph{What are the conceptual properties of DTs?}, \emph{How are 
DTs engineered, deployed, and operated?}, and \emph{To which extent are DTs evaluated?}\emph{What are the 
conceptual properties of DTs?}, \emph{How are DTs engineered, 
deployed, and operated?}, and \emph{To which extent are DTs evaluated?} These questions 
are addressed using both a vertical (quantitative) analysis, and an orthogonal analysis, which examines 
correlations among the outcomes of the vertical analysis. Their work serves as a crucial source 
of evidence for our own mapping study (cf.~\cref{sec:methodology}), as it offers a comprehensive analysis of 
various aspects related to DTs. This includes the types of research conducted and contributions made for DTs, 
the expected properties that DDTs must possess, the methods used to implement and operate DTs, and the evaluation 
process for DTs~\cite{dalibor2022cross}. This analysis is detailed and informative, making it valuable for 
proposing a generalized feature model for DTs. However, their approach does not address the unique challenge of 
developing a feature model for DTs.

Wagg et al.~propose a DT maturity model for synthesizing DTs which comprises five maturity levels 
(cf.~\cref{ssec:design-space})~\cite{wagg2020digital}. They evaluate their proposal by demonstrating 
how a DT synthesis process might be applied in practice~\cite{wagg2020digital}. Specifically, by illustrating 
modeling issues arising in the construction of a DT for a small-scale three-story structure 
they both highlight key processes including system identification, data-augmented modeling and verification 
and validation, as well as open issues and challenges. Their study offers a valuable perspective on a 
hierarchy of capabilities for DTs in terms of their level of development, which ultimately enables the 
identification of application domains and different types of twin functionality to be reflected in a \emph{generalized 
feature model} for DTs.

Su et al.~recently introduced a novel maturity model for DTs inspired by the Gemini-Principles, viz.,~purpose, 
trust, and function to enable a systematic view on the development and implementation of DTs~\cite{su13158224}. 
Following, these principles,  their maturity model comprises nine sub-dimensions, capitalizing 
on an expert questionnaire to develop 27 additional rubrics as part of the resulting maturity model. The model 
is eventually verified in two construction projects in Shanghai and Cambridge to qualitatively evaluate and 
compare the maturity of DTs in asset management. Su et al.'s work highlights the significance of considering 
socio-economic factors in the analysis and evaluation of DTs, in addition to focusing solely on functional and 
technical aspects~\cite{shahat2021city}. In its claim for generality, our suggested \emph{generalized feature model} 
for DTs encompasses these elements.

Similarly, Jeong et al.~\cite{jeong2022digital} propose a five-stage maturity model for the evolution of DTs: (1) mirroring 
for duplicating physical objects into DTs; (2) monitoring for using DTs for real-time control and analysis; 
(3) modeling and simulation for optimizing physical objects based on simulations; (4) federation for interconnecting 
multiple DTs to optimize complex systems; and (5) autonomous by allowing DTs to autonomously recognize and 
solve problems. Additionally, the paper outlines practical implementation layers, detailing specific technology
elements required at each stage to facilitate the development and integration of DTs across various industries, 
including energy, transportation, and smart cities. Contrary to our work, Jeong et al.'s study is neither backed 
by an SLR nor by practitioner feedback to strengthen and validate the results.

Apart from these rather domain-independent activities, various initiatives focused on the domain level,
as discussed in the following. Chen et al.~\cite{chen2024multiobjective} developed a comprehensive 
maturity model for building DTs in the construction industry proposing a framework that incorporates 
a fairness-aware multi-objective optimization approach to assess the maturity of building DTs. Their work 
includes the creation of a maturity model with seven dimensions and 22 indicators, and 
an evaluation method based on large-scale collective opinion generation. This framework is validated through 
a case study and aims to standardize maturity assessments, providing valuable insights for construction practitioners 
to enhance the development and diffusion of DT technology. By focusing only on the construction domain, their 
work, however, does not account for the larger context of DTs in other domains.

Uhlenkamp et al.~\cite{uhlenkamp2022digital} propose a maturity model specifically designed for the classification 
and evaluation of DTs within production and logistics systems. They highlight the diverse possibilities associated 
with DTs, emphasizing the need for formalization and standardization in their description and evaluation throughout 
their life cycle. Their maturity model assesses DTs across seven categories, viz.~context, data, computing 
capabilities, model, integration, control, and human-machine interface—comprising 31 ranked characteristics. 
The proposed model provides a systematic approach for evaluating existing DT solutions, identifying areas for 
improvement, and facilitating the transfer of DTs to new use cases. The model was validated through five use cases. 

In the course of their work on DTs for metal additive manufacturing, Phua et al.~\cite{phua2022digital} propose a 
DT hierarchy, organizing the development of DTs into four maturity 
levels: \emph{implicit}, \emph{instantiated}, \emph{interfaced}, and \emph{intelligent}.  Implicit DTs represent 
a basic virtual model based on existing knowledge of the manufacturing process, providing predictions without 
real-time feedback. Instantiated DTs enhance this by incorporating sensor data, creating a real-time virtual 
representation. Interfaced DTs integrate control mechanisms, allowing the system to adapt and optimize during the 
manufacturing process. Finally, intelligent DTs leverage advanced machine learning and AI for autonomous decision-making 
and continuous improvement. This hierarchical framework serves as a guide for developing DTs that address the 
complexities of metal additive manufacturing, with the ultimate goal of improving part qualification, 
certification, and process optimization.

Bellavista et al.~\cite{bellavista2024odte} recently introduced the \emph{Overall Digital Twin Entanglement} 
(ODTE) metric, a novel and concise measure of the synchronization quality between a DT and its physical twin. 
The ODTE metric focuses on two key factors: timeliness (freshness of the data) and completeness (ratio of collected 
to total data), providing a single normalized value that indicates the quality of the entanglement of the DT 
and the physical twin. This metric is crucial for ensuring that DTs accurately reflect and respond to the 
states of their physical counterparts, which is vital for their effective deployment in Industrial IoT 
environments. The paper also discusses the design principles for entanglement-aware DTs and validates the 
ODTE metric through practical experiments in a Kubernetes-based industrial test bed. The ODTE metric contributes 
to the maturity assessment of DT systems by offering a clear, application-agnostic way to monitor and 
improve DT performance, ensuring higher fidelity and reliability in real-world applications. Contrary 
to our work, Bellavista et al., however, ~do not investigate specific features, but rather provide 
ODTE as a tool for maturity assessment.

van der Valk et al.~\cite{van2022archetypes} present a taxonomy and classification of DTs based on a 
structured literature review and qualitative industry interviews. It identifies key characteristics and 
archetypes of DTs, emphasizing the variations in definitions and use cases across different industries. The 
study highlights the lack of standardization in DT concepts and proposes five archetypes that range from basic 
models to comprehensive, fully interoperable twins with autonomous control capabilities. Yet, their work 
does not provide a formalized feature model that explicitly defines dependencies, constraints, and hierarchical 
relationships between DT characteristics. While it classifies DTs based on clusters of characteristics, 
it lacks a structured framework to represent these features in a way that would facilitate systematic 
design or configuration selection.

Recently, Lehner et al.~\cite{lehner2025model} presented a systematic mapping study that investigates the 
application of MDE techniques for DTs across diverse domains. Their work identifies common patterns, challenges, 
and research gaps in leveraging MDE for automating aspects of DT engineering. Our work complements these findings 
by proposing a GFM for DTs, aiming to formalize and structure the mandatory and optional features of DTs to further 
support MDE workflows, design decisions, and verification processes.

Our discussion of related work reveals two important aspects. First and foremost, research around the concept 
of DTs is getting significant attention. However, over the last years, the general trend in engineering DTs was 
to build common-of-the-shelf, ad-hoc solutions. This has led to ignoring the underlying epistemology of DTs by properly 
characterizing DTs to understand what differentiates them from traditional executable models only. Second, 
and as we conjecture, this lack of complete understanding of DTs has resulted in various proposals for feature 
collections and maturity assessment. However, these proposals unequivocally emerge out of the research domain, 
which provides only one concrete view of DTs. 

To fully understand DTs and successfully harvest their potential and benefits, a clear \emph{domain-independent} 
understanding of the concept and epistemology of DTs is necessary, which comprises a feature model that clearly 
outlines the concept of DTs in terms of relevant functionality. This is the goal of our work.

\section{Methodology}
\label{sec:methodology} %former mapping-study

The GFM development adheres to the DSR process model proposed by Wieringa~\cite{wieringa_dsr_2014} which 
is widely referenced and recognized for performing design science research, to design and develop artifacts that solve identified problems. 

Following this, a systematic literature review (SLR) was conducted to compile DT features mentioned in 
scientific literature and gain an understanding of their interrelationships and context. Consequently, the 
SLR serves as the basis for the subsequent development of the GFM, acting as its primary source of information. Second, 
in the context of feature modeling, the features identified through the SLR underwent a rigorous and iterative 
process of coding and categorization, merging, and structuring, resulting in a cross-domain GFM of a DT. This 
process was repeated until most of the features identified in the SLR were categorized, the derived features 
were clearly distinguishable by name and content, and the GFM followed established feature modeling notations 
while accurately representing the features and their relationships. Based on the GFM of the DT, in addition, 
two distinct versions of the feature model were 
then derived, one representing the \emph{digital shadow} (DS) and the other the \emph{digital model} (DM).

The three versions of the GFM provide a comprehensive overview of the derived DT features, including their 
relations and varying levels of necessity, thus enabling the distinction between digital models, shadows, 
and twins. To further refine the GFMs and to illustrate their universality, domain-specific aspects and 
differences among the derived features were highlighted. Readers are encouraged to engage with the GFMs 
or conduct further research on them, helping to refine and enhance them through feedback.

\subsection{Systematic Literature Review}
\label{subsec:slr}
The SLR aligns with the Preferred Reporting Items for Systematic Reviews and Meta-Analyses (PRISMA) 2020 
statement outlined by Matthew et al.~\cite{page2021prisma-a}. This ensures a systematic method for 
conducting reviews, where PRISMA 2020 serves as an updated reporting guide with advanced 
methodologies. PRISMA, originally designed for systematic reviews in the health sector, 
has been utilized in various other disciplines~\cite{page2021prisma-a,page2021prisma-b}. For example, 
Uhlenkamp et al.~\cite{uhlenkamp2022digital}, employed the PRISMA statement for an SLR to create 
a maturity model for DTs (cf.~\cref{sec:background}). 

The current SLR synthesizes the characteristics of DTs as presented in the scientific literature across 
key application domains, establishing a foundation for the development of three versions of the cross-domain 
GFM. The focus is exclusively on the characteristics of DTs, excluding DSs and DMs. The DT is more 
extensively documented in the literature, whereas a DT includes all the features of a DS and a DM.

\subsubsection{Eligibility Criteria}

The literature examined in the SLR is constrained by a set of inclusion and exclusion criteria, as 
presented in Tbl.~\ref{tbl:criteria}, to improve the relevance and quality of the publications analyzed 
in relation to the desired outcomes. To ensure readability and accessibility, the SLR included only 
literature published in English that provided full-text access. Furthermore, only literature from 
scientific journals and conference proceedings was included, while editorials and perspective articles 
were excluded due to their potential subjectivity. Books, conference abstracts, and publications with 
a total length of fewer than six pages, including the abstract, appendix, and references, were excluded 
due to insufficient detail. Additionally, literature identified in Beall’s list of potential predatory 
journals and publishers was also excluded~\cite{beall2012predatory}. 
Given the rapid evolution of the concept of DTs~\cite{sharma2022digital}, it is pertinent to limit 
the literature review to a more recent timeframe.

The findings of Uhlenkamp et al.~\cite{uhlenkamp2022digital} support this assertion, revealing that 
the inaugural literature review on DTs appeared in 2015, with a notable increase in both the number 
of reviews and references per review occurring in 2018. Consequently, in alignment with the reviews 
conducted by Botín-Sanabria et al.~\cite{botin2022digital} and Thelen et al.~\cite{thelen2022comprehensive}, 
the SLR incorporated only the literature published from 2017 to May 2024. 

\begin{table*}[bth]
\small
\caption{Inclusion and Exclusion Criteria of the SLR}
\label{tbl:criteria}
\renewcommand{\arraystretch}{1.15}
\begin{tabular}{@{}p{0.48\textwidth} p{0.48\textwidth}@{}}
\toprule
\textbf{Inclusion Criteria} & \textbf{Exclusion Criteria} \\ \midrule
Literature in English & \\
Literature from scientific journals and conference proceedings & Books, conference abstracts, perspective articles, and editorials \\
Literature with full-text accessibility & Literature listed in Beall’s list of potential predatory journals and publishers \\
Literature published between 2017 and May 2024 & Literature shorter than six pages (including abstract, appendix, and references) \\
Focus on the concept of Digital Twins (DTs) & Focus on concepts where DTs are not the core topic \\
Literature that contains features of DTs & \\
Literature related to at least one application domain: manufacturing, smart city and urban, healthcare, automotive, aerospace, maritime & \\ 
\bottomrule
\end{tabular}
\end{table*}

The literature needed to concentrate on the concept of DTs to be deemed relevant. This encompasses 
variations of DT, including cognitive DTs and DT-based systems. Literature pertaining to concepts 
linked to DTs, where DTs are not central to the concept, was excluded. Another inclusion criterion 
was that the literature must include features of DTs. This may also encompass attributes referenced 
by authors. 

The literature needed to pertain to at least one of the predefined application domains to be included 
in the SLR. This guarantees a transparent distribution of the DT features to an application domain, a 
precise delineation of the proposed features in relation to the addressed application domains, and that 
the resulting feature models are applicable across domains.  Publications addressing DTs at a domain-agnostic level were included, as they are relevant across various application domains and frequently discuss fundamental and broadly applicable characteristics. Focusing solely on domain-specific literature may result in the neglect of fundamental features of DTs.

Identifying the most relevant and commonly represented application domains of DTs in scientific 
literature was essential for determining the predefined application domains. The process entailed 
the analysis of five published reviews and surveys concerning DTs. Despite variations in search 
strategies and analytical methods, three publications produced comparable results. Uhlenkamp et al.~\cite{uhlenkamp2022digital} 
identify the five primary application domains with the highest number of published literature 
reviews on DT up to 2021 as manufacturing, aerospace, healthcare, automotive, and maritime. Smart 
cities rank seventh~\cite{uhlenkamp2022digital}. Barricelli et al.~\cite{barricelli2019survey} 
identified manufacturing, aviation, and healthcare as the main application domains of DTs. Semeraro~\cite{semeraro2021digital} 
identified five primary application domains: manufacturing, aerospace, maritime and shipping, 
healthcare, and city management. The findings were further supported by the reviews conducted 
by Botín-Sanabria et al.~\cite{botin2022digital} and Thelen et al.~\cite{thelen2022comprehensive}. 
This survey identifies manufacturing, aerospace/aviation, automotive, healthcare/medicine, 
maritime/marine/shipping, and urban environments as the primary application domains of DTs.

\subsubsection{Information Sources and Search Strategy}

Literature was gathered through searches of the databases Web of Science\footnote{\url{https://www.webofscience.com/wos}}, 
ScienceDirect\footnote{\url{https://www.sciencedirect.com}}, IEEE Xplore\footnote{\url{https://ieeexplore.ieee.org}}, 
and EBSCOhost\footnote{\url{www.ebsco.com}}. Tbl~\ref{tbl:results} presents the search date, utilized search strings, 
applied search fields and filters, along with the number of records identified.

The SLR seeks to integrate the characteristics of DTs within the most pertinent application domains. A 
search string was formulated and segmented into three components to accomplish this objective. The initial 
section addresses DTs, referred to as \SSr{"Digital Twin*"}. The second part addresses the characteristics 
of DTs. Autiosalo et al.~\cite{autiosalo2019feature} assert that a DT feature denotes the technical functionalities 
of a DT. Consequently, searches were conducted using the terms \SSr{"Feature*"} and \SSr{"Functionalit*"}. 
The third component of the search string includes the various application domains to be incorporated into 
the SLR. The domains, terms, and synonyms were chosen based on findings and terminology from five prior reviews 
and surveys~\cite{barricelli2019survey,botin2022digital,semeraro2021digital,thelen2022comprehensive,uhlenkamp2022digital}. 
This led to the emergence of the terms manufacturing, aerospace, aviation, automotive, healthcare, medicine, 
city, cities, urban, maritime, shipping, and marine. In each of the three sections, the presence of at least 
one word or phrase from the search string was necessary for an article to meet the search criteria. A 
limitation to particular application domains and terminology is essential due to the absence of a universally 
applicable list of application domains for digital twins. Otherwise, clearly and transparently assigning 
publications to specific application domains would be impossible.
\begin{table*}[bth]
\small
\centering
\begin{threeparttable}
\caption{Search Overview: Databases, Dates, Queries, Fields, Filters, and Records}
\label{tbl:results}
\renewcommand{\arraystretch}{1.15}
\begin{tabular}{@{}p{2cm} p{1.4cm} p{5.2cm} p{2.5cm} p{2.7cm} p{0.8cm}@{}}
\toprule
\textbf{Database} & \textbf{Date} & \textbf{Search String (simplified)} & \textbf{Search Fields} & \textbf{Filters} & \textbf{Records} \\ \midrule

Web of Science & May 3, 2024 & 
\texttt{TI/AK=("Digital Twin*") AND TS=(Feature* OR Functionalit*) AND TS=(App. Domains)} & 
Title, Abstract, Keywords, Topic & 
Article and Review Article & 243 \\

ScienceDirect & May 3, 2024 & 
\texttt{Title/Abstract/Keywords: ("Digital Twin") AND (Feature OR Functionalities) AND (App. Domains)} & 
Title, Abstract, Keywords & 
Review Article and Research Article & 126 \\

IEEE Xplore & May 3, 2024 & 
\texttt{Title/AK/IEEE Terms: "Digital Twin*" AND All Metadata: (Feature* OR Functionalit*) AND (App. Domains)} & 
Title, Author Keywords, IEEE Terms, All Metadata & 
Conferences, Journals & 281 \\

EBSCOhost & May 3, 2024 & 
\texttt{TI/SU "Digital Twin*" AND (Feature* OR Functionalit*) AND (App. Domains)} & 
Title or Subject Terms & 
Boolean mode; Peer-reviewed; Academic journals & 111 \\

\bottomrule
\end{tabular}

\vspace{0.5em}
\begin{tablenotes}
\item \textbf{Abbreviations:}
\item \textbf{TI} = Title; \textbf{AK} = Author Keywords; \textbf{TS} = Topic Search; \textbf{SU} = Subject Terms
\item \textbf{App. Domains} = Application domains: Manufacturing, Aerospace, Aviation, Automotive, Healthcare, Medicine, City, Cities, Urban, Maritime, Shipping, Marine
\item \textbf{—} = Not reported or not applicable
\end{tablenotes}
\end{threeparttable}
\end{table*}

To ensure that publications focus on DTs, the phrase "Digital Twin" must be included in the title or 
keywords of the publications in the majority of databases. To enhance the quality and relevance of the publications, all other 
terms were primarily examined in the title, keywords, and abstract. The filters utilized in the databases adhere to the specified 
inclusion and exclusion criteria (cf.~Tbl.~\ref{tbl:criteria}).

\subsubsection{Selection Process}

The initial step involved the de-duplication of records identified within the databases. The titles and abstracts 
of the records were subsequently evaluated according to the predetermined inclusion and exclusion criteria. 
Inappropriate records were excluded, and full-text access was confirmed for the remaining records. The full 
texts of the remaining publications were evaluated according to the established inclusion and exclusion criteria.

In the full-text screening stage, the inclusion criteria for DT features were further delineated. Alongside 
features, functionalities were also examined, as Autiosalo et al.~\cite{autiosalo2019feature} indicate that 
a DT feature denotes the technical functionalities of a DT. The two terms characterize the utilized search 
string. To mitigate the risk of overlooking features, relevant characteristics, properties, and attributes of 
DTs were also examined in the screened full-texts, as these terms are frequently used interchangeably 
with the term feature in the literature. Due to the generality of the term "feature," which lacks precise definition 
and boundaries, it is necessary for features, functionalities, characteristics, properties, and attributes to 
be explicitly labeled in publications or to be evident from the context to be considered. This study generally 
categorizes these terms as features. To avoid distortion of the resulting feature models, aspects such as components, 
services, functions, and applications have been excluded, despite occasional overlaps and labeling as features, 
as these typically represent different dimensions of the digital twin that exceed the scope of the SLR. Furthermore, 
features exclusively associated with DT constructs, such as architecture, were omitted for the same rationale. 
In conclusion, these inclusion details seek to provide a comprehensive overview of the features of DTs while 
remaining within the thesis's scope and ensuring transparency with minimal bias.
\begin{figure*}[bth]
\centering
    \includegraphics[width=.7\textwidth]{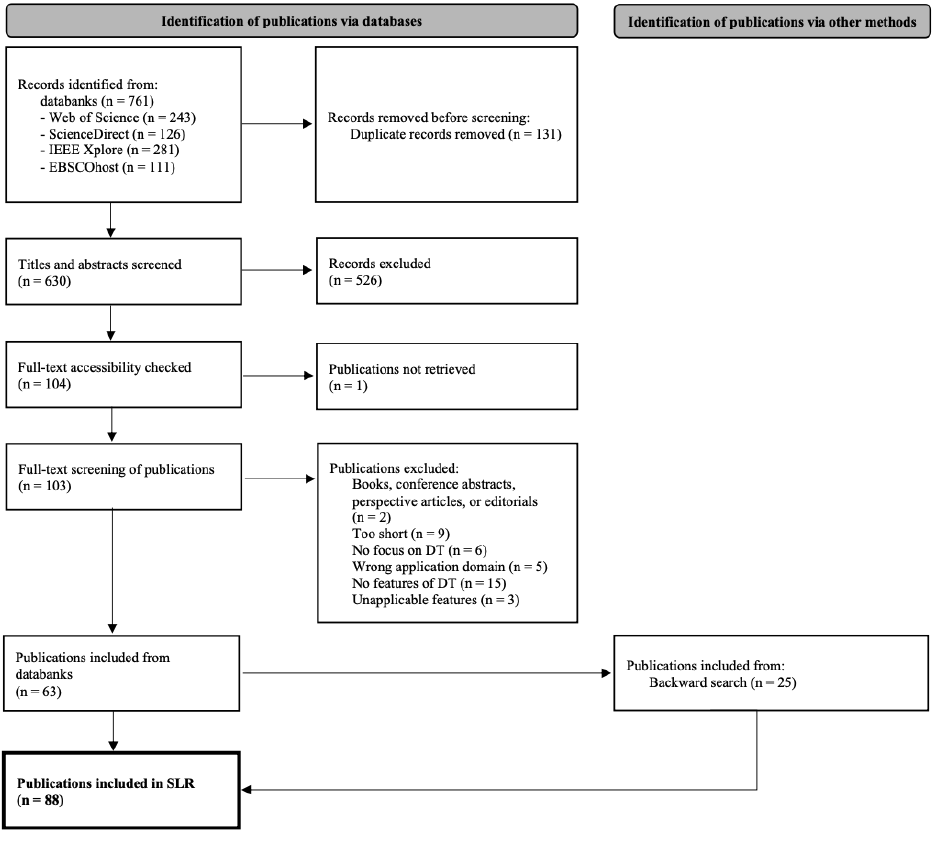}
    \caption{PRISMA Flow Diagram~\cite{page2021prisma-a}.}
    \label{fig:search}
\end{figure*}

Full-text screening identified six criteria for the exclusion of publications, as detailed in Fig.~\ref{fig:search}. 
All other publications were incorporated into the systematic literature review. A reverse search was performed on 
the remaining publications by reviewing and evaluating their references using the same selection criteria. In total, 
63 publications were identified through databases, and 25 additional publications were found via backward search, culminating 
in 88 publications included in the SLR. The selection process was conducted manually, without employing automation 
tools. The PRISMA flow diagram (cf.~Fig.~\ref{fig:search}) illustrates all results and the steps undertaken.

\subsection{Feature Modeling}
\label{ssec:feature-modeling}

The feature models were developed using the feature-oriented domain analysis (FODA) method as proposed by 
Kang et al.~\cite{kang19904a1}, which established the foundational concept of feature modeling~\cite{van2023reference,benavides2010automated}. 
The objective is to identify standard and prevalent characteristics of software products within a specific 
domain~\cite{apel2009overview,kang19904a1}. FODA comprises three phases: context analysis, domain modeling, 
and architectural modeling, with feature modeling being a fundamental component~\cite{apel2009overview,kang19904a1}. 
Domain modeling consists of three procedures: feature analysis, entity-relationship modeling, and 
functional analysis~\cite{kang19904a1}. In this article, we primarily focus on developing a feature model and 
in turn context and feature analysis within domain modeling.

The key elements of the context analysis, such as domain coverage and delimitation, scope, and utilized 
information sources, are thoroughly discussed in the SLR and elaborated upon in Section~\cref{subsec:slr}~\cite{kang19904a1}.

During the feature analysis and domain modeling phase, features were first identified via the SLR. These 
features underwent a systematic process of coding, categorization, merging, and structuring, culminating 
in a cross-domain hierarchical feature model of a DT. The process was iterated to ensure that as many features 
identified in the context analysis as possible were considered and categorized. The derived features were 
distinguishable in both name and content at an adequate level, and the feature model conformed to established 
feature modeling notations while accurately representing the features and their interrelations. Two distinct 
versions of the feature model were derived from the feature model of the DT, one representing the DS and the 
other the DM.

\subsubsection{Determination of Possible, Optional, and Mandatory}

The definitions of parent, child, concrete, and abstract features, along with their interrelationships, were 
established through the SLR. Numerous definitions and interpretations of Digital Twins and their necessary 
and potential characteristics are present in the literature~\cite{kritzinger_digital_2018,semeraro2021digital}. 
Furthermore, numerous concepts of DTs fail to distinguish between a digital model, shadow, and twin, 
thereby complicating the comprehension of digital twins~\cite{kritzinger_digital_2018}. Classifying features 
into mandatory and optional categories concerning digital models, shadows, and twins presents a challenging 
and inconsistent process. The identification of possible, optional, and mandatory features is primarily focused 
on differentiating and defining digital models, shadows, and twins based on their integration levels, as 
outlined by Kritzinger et al.~\cite{kritzinger_digital_2018}. The levels of integration proposed by Kritzinger et 
 al.~\cite{kritzinger_digital_2018} are well-suited for developing the three versions of the feature model, as 
 they address the concept of all three digital twin types from a foundational perspective and highlight the key 
 distinctions among them.  The classification of all features cannot be established solely based on these levels; 
 therefore, information from the literature regarding the features was also considered, necessitating justified 
 assumptions where appropriate.

Features are categorized as mandatory if they are essential components of every DT, DS, or DM, respectively. 
Alternatively, they are classified as optional. Therefore, features that are typical or commonly observed in a 
DT do not inherently meet the criteria to be classified as mandatory features. It is important to recognize 
that a feature may exhibit varying levels of functionality and maturity.  Thus, a feature classified as mandatory 
indicates that it must be present in every DT at a fundamental level.

\subsubsection{Exclusion of Features}
\label{sssec:exclusion}

Some features identified by the SLR were not included in the resulting feature models due to at least one of 
four reasons: (1) They could not be categorized due to insufficient context and significant room for interpretation; 
(2) They are not seen as part of a DT:  (3) They do not fit in the feature models due to their hierarchical level 
or granularity; (4) They are only mentioned once in the literature and do not contribute value to the feature 
models. The fourth reason is based on maintaining the feature models as complex as necessary and as simple as 
possible.
\section{A Generalized Feature Model for Digital Twins}
\label{sec:models}

In this section, we first present the quantitative results in~\cref{ssec:quantitative} and in~\cref{ssec:features} we elaborate on and discuss the derived features, and the GFM, including the 
distinction between the DM, DS, and DT. Finally, in~\cref{ssec:domains} we provide a domain-specific view of the 
features to highlight differences between the domains. Accordingly, the results section is structured in a way 
that begins with a cross-domain overview and then becomes increasingly 
specific throughout~\cref{ssec:domains}.

\subsection{Quantitative Results}
\label{ssec:quantitative}

\begin{figure}[bth]
\centering
    \includegraphics[width=\columnwidth]{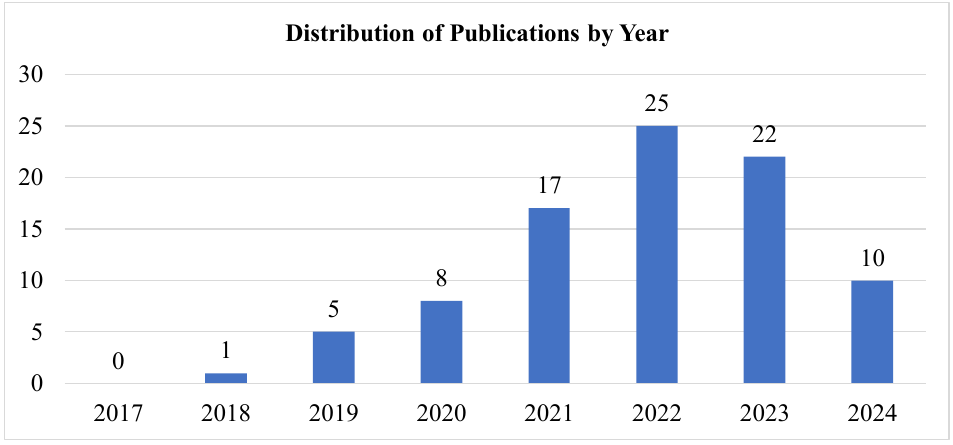}
    \caption{Number of Publications on DT Features in Relevant Application Domains (2017 to May 2024).}
    \label{fig:pubs-per-year}
\end{figure}

\begin{figure}[bth]
\centering
    \includegraphics[width=\columnwidth]{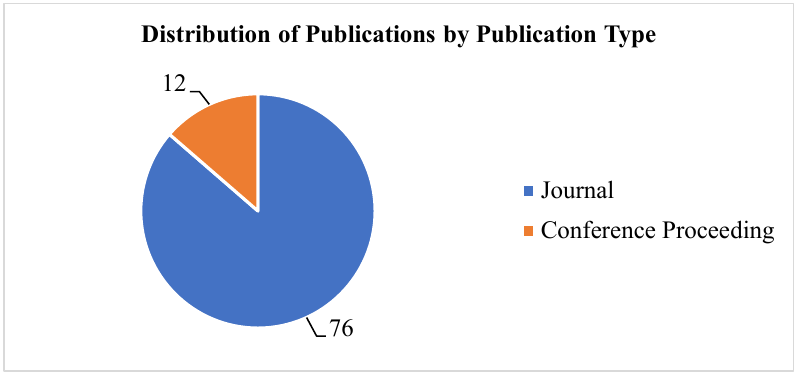}
    \caption{Number of Publication on DT features in Relevant Application Domains by Publication Type.}
    \label{fig:pubs-dist}
\end{figure}

Fig.~\ref{fig:pubs-per-year} confirms that the number of publications on DT features within relevant application 
domains has increased consistently in recent years, peaking in 2022 and 2023. However, 2024 can only be assessed 
to a limited degree, as the literature search encompassed only the initial four months of the year, corresponding 
to the execution date of the initial literature search. The number of publications for 2024 has already surpassed 
the total from 2017, 2018, and 2019 combined. These findings further reinforce the growing significance and interest in DTs 
in academic research. The majority of these publications are found in journals, with only about 14\% appearing 
in conference proceedings (cf. Fig.~\ref{fig:pubs-dist}).

\begin{figure}[bth]
\centering
    \includegraphics[width=\columnwidth]{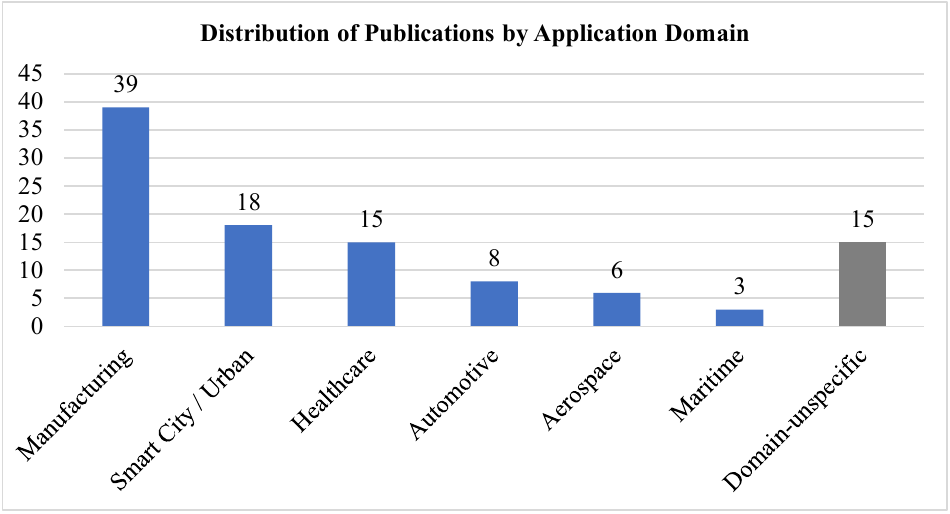}
    \caption{Number of Publications on DT Features per Application Domain.}
    \label{fig:pubs-domain}
\end{figure}

The distribution of publications across application domains, as illustrated in Fig.~\ref{fig:pubs-domain}, indicates 
that the majority of DT literature concerning features is either domain-agnostic or pertains to manufacturing (39 publications), 
smart city (18 publications), or healthcare (15 publications). The SLR and the feature models developed in 
the subsequent section are primarily influenced by the manufacturing domain, with secondary influences from the smart 
city and healthcare sectors. The maritime and aerospace sectors (6 and 3 publications respectively) are underrepresented 
in the feature models due to a limited number of suitable publications. The domain-agnostic publications encompass various 
application domains, thereby enhancing the general applicability and completeness of the resultant feature models. Fig.~\ref{fig:pubs-domain} 
indicates that the total number of publications presented surpasses the total number of publications incorporated in the 
systematic literature review, as certain publications pertain to multiple application domains.

\begin{figure}[bth]
\centering
    \includegraphics[width=\columnwidth]{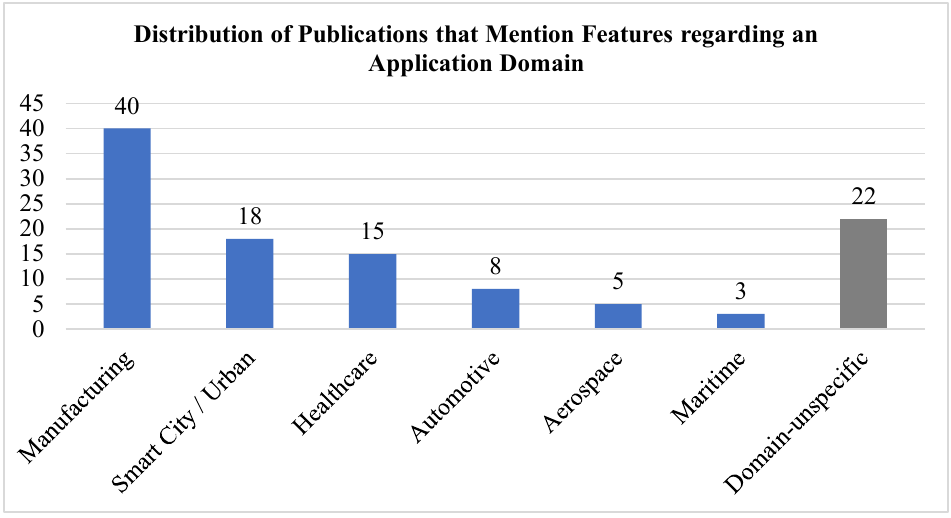}
    \caption{Number of Publications on DT Features that Mention Features Regarding an Application Domain.}
    \label{fig:pubs-dom-feature}
\end{figure}

Nonetheless, the reference of a publication to a specific application domain does not imply that all DT features 
discussed are inherently linked to that domain. Consequently, the subsequent process was executed: Features are allocated 
to domains according to the application domain or described decision trees from the relevant publications. Thus, 
the assignment of a feature to a specific application domain does not preclude its applicability to DTs 
from other domains. However, if it was clearly stated that the mentioned features do not pertain to the application 
domain or the presented DT of the publication, but rather to another domain or DTs at a 
domain-agnostic level, this was taken into account. Consequently, Fig.~\ref{fig:pubs-dom-feature} illustrates 
the number of publications referencing DT features across various application domains.

Fig.~\ref{fig:pubs-dom-feature} indicates no differences among the smart city, healthcare, and maritime domains in 
comparison to Fig.~\ref{fig:pubs-domain}. One publication in the aerospace domain did not reference any DT features 
relevant to this application area, whereas another publication, outside the manufacturing domain, emphasized DT 
features applicable to manufacturing. Additionally, various publications from particular application domains have 
referenced domain-agnostic DT features. The allocation of application domains to publications is now available 
from our Zenodo repository~\cite{zech_2025_15259964}.

\subsection{Derived Features and Feature Models}
\label{ssec:features}

\begin{figure}[bth]
\centering
    \includegraphics[width=\columnwidth]{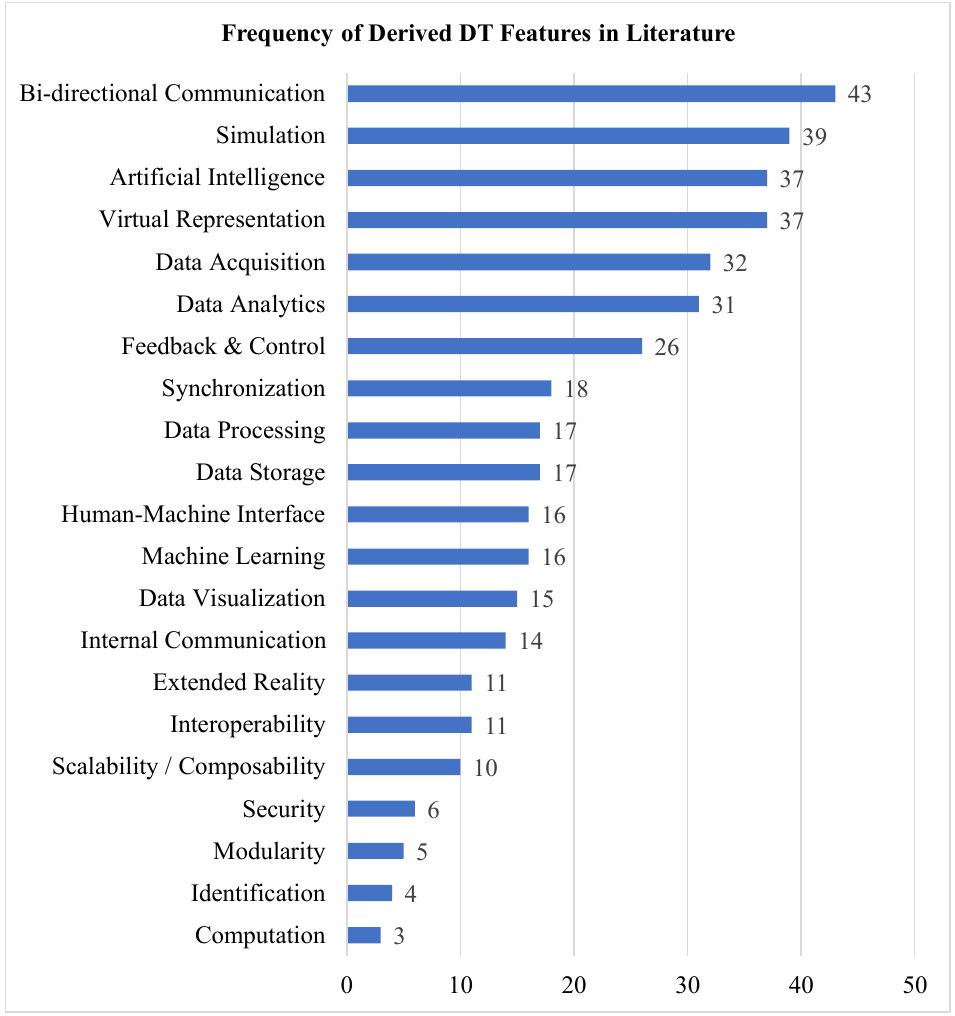}
    \caption{Frequency of 21 Comprehensive DT Features in Literature Derived from SLR: Arranged in Descending Order.}
    \label{fig:feature-count}
\end{figure}

The SLR identified several hundred design space features from 88 publications across six application domains, distilled into 
21 comprehensive features, as illustrated in Fig.~\ref{fig:feature-count}. The detailed allocation of derived DT features to publications 
is available from our Zenodo repository~\cite{zech_2025_15259964}. The reviewed literature identifies bi-directional communication, 
simulation, AI, and virtual representation as the most prominent features of DTs. Conversely, security, modularity, identification, 
and computation are the features least frequently mentioned. The prevalence of bi-directional communication as a 
predominant characteristic of a DT 
corresponds with the conceptualization of Digital Twins articulated by Kritzinger et al.~\cite{kritzinger_digital_2018}. 

Furthermore, the inclusion of simulation and virtual representation as frequently mentioned features aligns with the prevalent definitions 
of DTs~\cite{kritzinger_digital_2018,tao2018digital}. 

%The derived DT features align with existing research literature.

Commonly, a feature consists of multiple functionalities and varying maturity levels. Data processing may 
include basic tasks such as data filtering and aggregation, as well as more complex activities like data fusion and development of ontologies~\cite{barricelli2019survey,popescu2022building,de2022digital}. Synchronization may be executed 
either iteratively, on an hourly basis, or in real-time based on events~\cite{uhlenkamp2022digital,li2022digital}. Additionally, 
it is important to recognize that the derived features may not correspond to the same hierarchical level. Computation serves 
as a fundamental element underlying various features, including data analytics and simulations~\cite{autiosalo2019feature}. 
Artificial intelligence can represent multiple hierarchical levels, enhanced by simulations and data analytics; it may also 
facilitate control mechanisms~\cite{barricelli2019survey,lu2020digital}. The derived features are interdependent, influencing 
and facilitating one another, or constructed upon each other. It is important to note that the derived features and feature 
models encompass the majority of the several hundred identified features from the SLR, although not all, due to various 
reasons outlined in~\cref{sssec:exclusion}. Below, an example for each reason is provided.

Features related to manageability and ownership of the DT were omitted due to the fact that they were mentioned only once in the
literature and were considered standard functionality. Adding them, from our perspective, would have unnecessarily complicated the 
feature models without providing additional value~\cite{minerva2020digital}. 
In certain instances, the DT was characterized as integrative and open~\cite{jeddoub2023digital}. Nevertheless, the absence of 
contextual information regarding these features precluded categorization. The physical entity was frequently designated as 
a DT feature~\cite{jones2020characterising,vanderhorn2021digital,zheng2022emergence}. This work supports the perspective of 
Autiosalo et al.~\cite{autiosalo2019feature} and other researchers who view the physical entity as separate from the DT, rather 
than as a part of it. Consequently, the physical entity was excluded.

Some features were ultimately left out because of their unsuitable hierarchical level and granularity. The features identified 
in the SLR exhibited considerable variation regarding granularity and type. Limitations in feature modeling prevent the implementation 
of all feature types within a single model without introducing significant complications. Prediction, simulation, and data analytics 
cannot be treated as separate features within the same feature model, as either simulation or data analytics can facilitate prediction. 
Simultaneously, these two features may fulfill functions beyond mere prediction. Moreover, simulation and data analytics are distinct 
entities, each encompassing unique functionalities. Data analytics and simulations possess predictive capabilities, yet each also 
features distinct attributes that surpass mere predictions, rendering it challenging to effectively merge or categorize all three. 
Thus, during the three iterations of the feature model, a decision was required on whether to prioritize technical features 
or to focus on functional objectives. A technical focus distinguishes features like data analytics, simulations, machine 
learning (ML), and AI, while a purpose-oriented focus differentiates between monitoring, optimization, and prediction features. 
DT features are characterized as technical functionalities~\cite{autiosalo2019feature}. Given that features like monitoring and 
optimization can be facilitated through these technical aspects, we concluded that integrating features as technical enablers 
yields greater value than incorporating features based solely on functional purposes. Consequently, certain features identified in the 
SLR, such as monitoring and optimization, were incorporated into the feature models only when a definitive allocation to one or more 
technical features was achievable. Additional features that were excluded for similar reasons include validation, collision detection, 
data, and testing~\cite{hassani2022impactful,loaiza2023proposing,wallner2023digital}.

Furthermore, the decision was made to exclude feature requirements represented as cross-tree constraints in the feature models. This decision 
arises from the significant interconnectivity and the differing levels of maturity and functionality of the features, as described above. Consequently, 
various features may necessitate numerous additional features, and the required features may differ based on functionality and maturity 
level. The literature frequently lacks adequate information regarding the specific requirements of features. Clear and essential 
requirements are represented in feature models by effectively utilizing parent and child features, along with optional and mandatory 
features. This approach is suitable as child features inherently necessitate parent features, and all mandatory features are uniformly 
implemented; thus, their requirement status is irrelevant. Utilizing various relationships among features for modeling requirements 
is valid and can often be the preferred approach. This is because introducing constraints may result in redundancies concerning the 
model's semantic information~\cite{benavides2010automated}. Additional potential requirements of the derived features, which are not 
apparent in the feature models, are discussed in detail below.

\begin{figure}[bth]
\centering
    \includegraphics[width=\columnwidth]{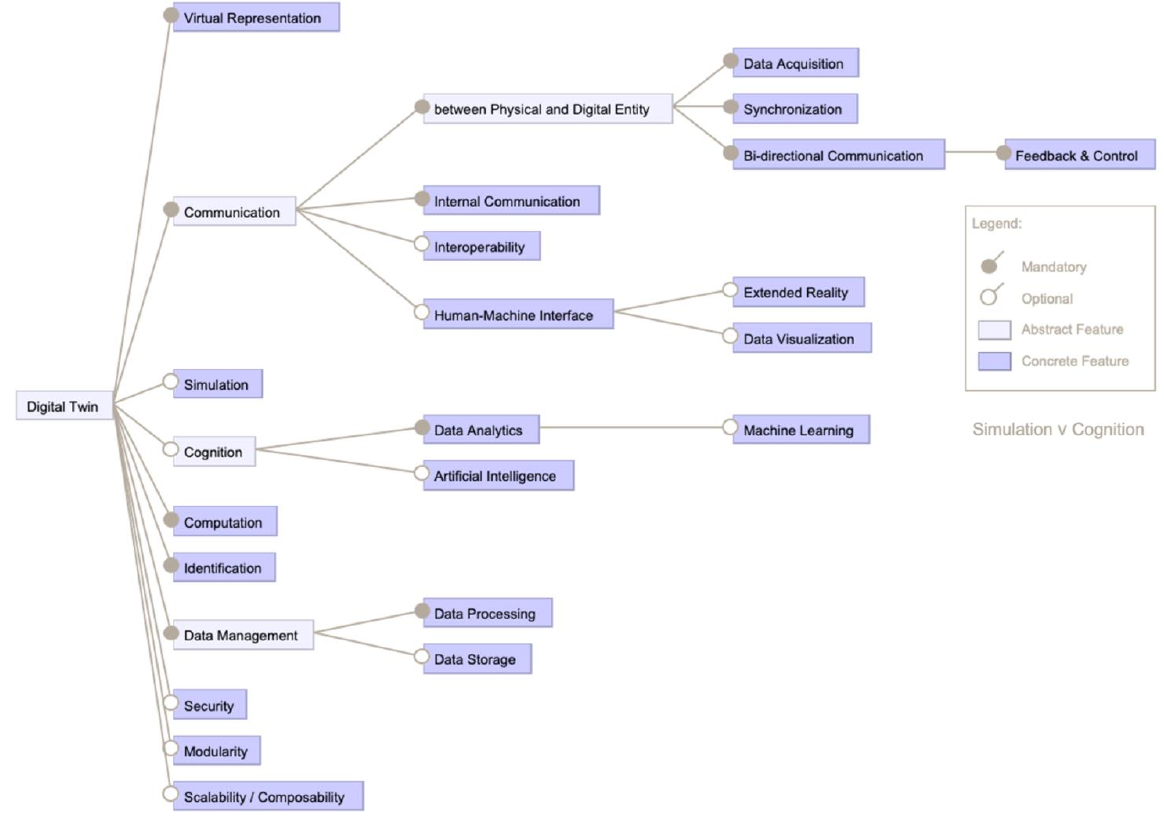}
    \caption{GFM of a DT: Encompassing all 21 Derived Features; Created with FeatureIDE (Eclipse).}
    \label{fig:dt}
\end{figure}

\begin{figure}[bth]
\centering
    \includegraphics[width=\columnwidth]{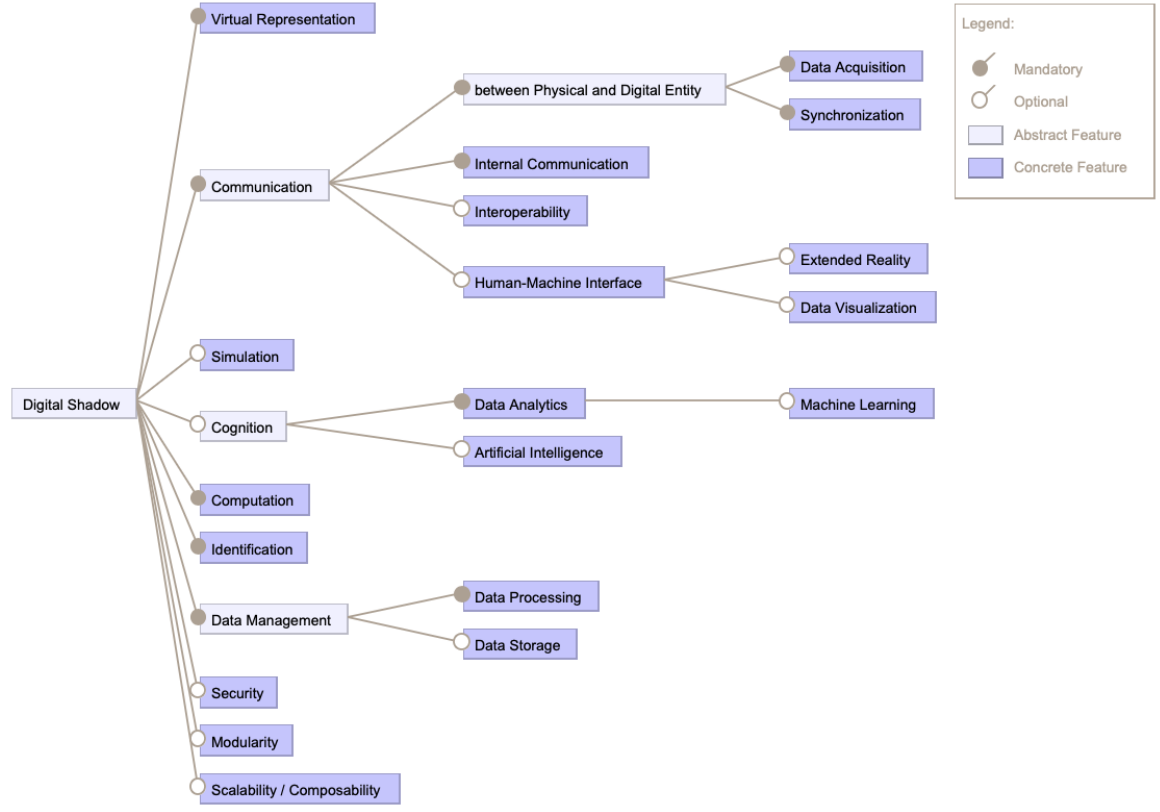}
    \caption{GFM of a DS: Derived from DT GFM; Encompassing 19 Derived Features; Created with FeatureIDE (Eclipse).}
    \label{fig:ds}
\end{figure}

\begin{figure}[bth]
\centering
    \includegraphics[width=\columnwidth]{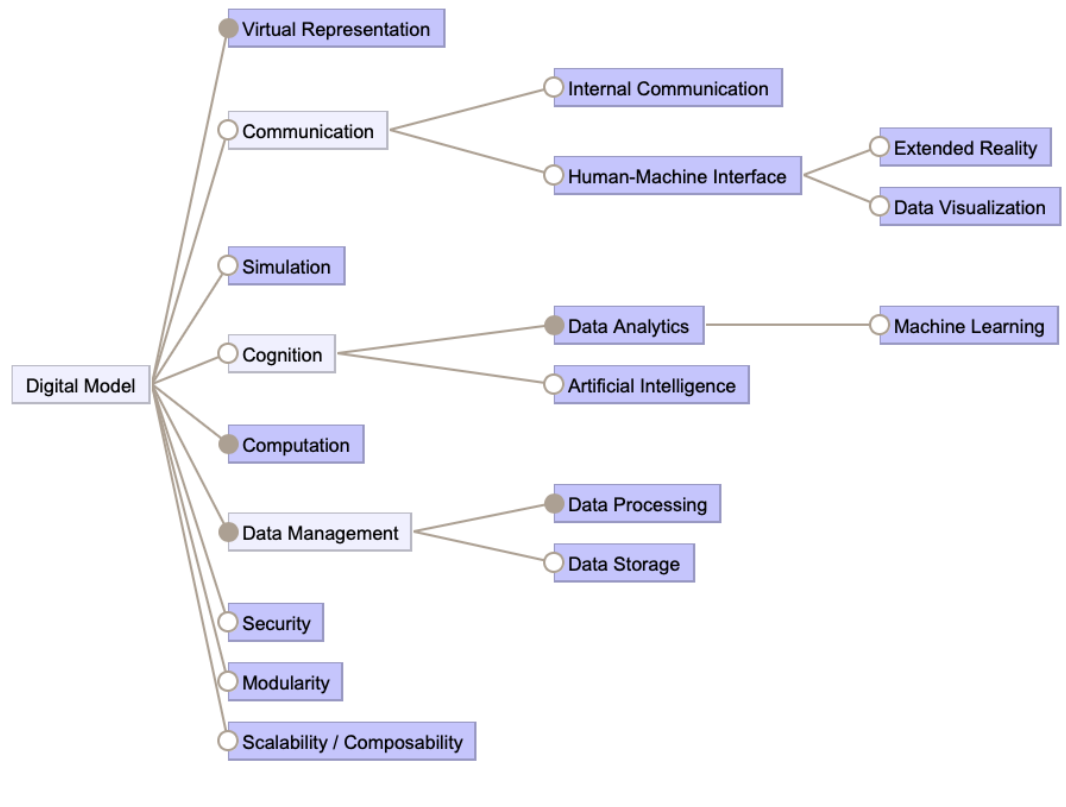}
    \caption{GFM of a DM: Derived from DT GFM; Encompassing 15 Derived Features; Created with FeatureIDE (Eclipse).}
    \label{fig:dm}
\end{figure}

Fig.~\ref{fig:dt} illustrates the cross-domain feature model of a DT, incorporating all derived features presented in Fig.~\ref{fig:feature-count} 
and highlighting their interrelations and significance. This indicates the essential features required in every decision tree and 
those that may be included. Features connected by a consist-of relationship denote the hierarchical relationship between parent 
and child features, resulting in content and functionality overlap. The features are distinct at an adequate level, while remaining 
interconnected. The feature model was developed using the processes and methods outlined in~\cref{ssec:feature-modeling}. The GFM 
of a DS (cf.~Fig.~\ref{fig:ds}) and a DM (cf.~Fig.~\ref{fig:dm}) were derived from the DT GFM, following the definition 
established by Kritzinger et al.~\cite{kritzinger_digital_2018}. The DT represents the highest level of integration, while DM 
represents the lowest among the three types. Feature models enhance the number and functionality of features from DM to DT and 
vice versa~\cite{kritzinger_digital_2018}.


\begin{figure}[bth]
\centering
    \includegraphics[width=\columnwidth]{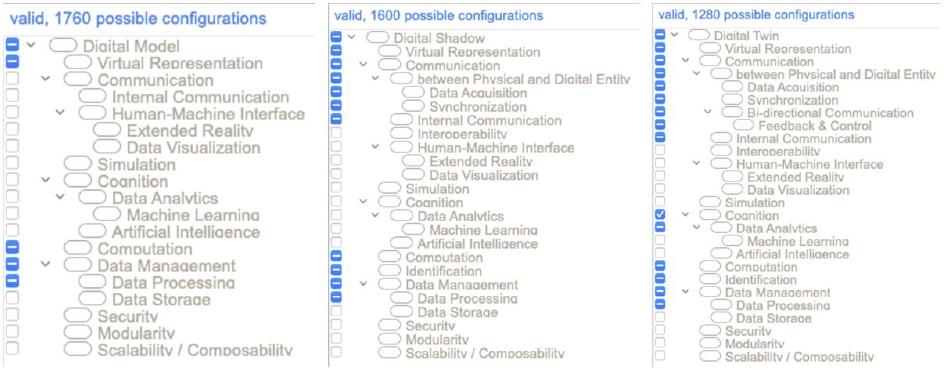}
    \caption{Validated Default Selection of Features for the DM, DS, and DT GFMs According to FeatureIDE: Includes the 
       Number of Possible Configurations; Created with FeatureIDE (Eclipse).}
    \label{fig:feature-validation}
\end{figure}

Fig.~\ref{fig:feature-validation} illustrates that the automatic validation process of FeatureIDE~\cite{kastner2009featureide} 
confirms the validity of all three versions of the feature model. The default selection of features for each decision tree type 
is clear, as is the number of potential configurations. The or-constraint in the DT feature model necessitates the manual 
selection of either the cognition feature, which denotes data analytics, or the simulation feature in the default configuration. 
Fig.~\ref{fig:feature-validation} further illustrates that cognition was selected; nonetheless, opting for simulation would have 
yielded an equivalent outcome.

\vspace{.25cm}
%\todo[inline]{Needs to be aligned with section 2, design space and solutions space, as well as concepts introduced there}

In the following, we provide a comprehensive description of all derived features part of the three proposed models. We further argue about the 
necessity and interrelationships of each feature.

\subsubsection{Virtual Representation}
The \emph{virtual representation} encapsulates the fundamental nature of each DM, DS, and DT~\cite{kritzinger_digital_2018}. This 
feature appears in almost half of the reviewed publications in the SLR and is a critical component of most, if not all, definitions 
of DTs~\cite{semeraro2021digital}. Virtual representation refers to the capacity to accurately and realistically mirror, reflect, 
replicate, model, or describe a physical entity depicting the real world. This representation must be comprehensive, 
encompassing all pertinent data and characteristics of the physical entity, and should exhibit high or multi-level fidelity. The 
represented physical entity includes the state, behavior, environment, shape, and additional information pertaining to the physical 
counterpart~\cite{hassani2022impactful,hribernik2021autonomous,jeddoub2023digital,jones2020characterising,li2022digital,loaiza2023proposing,
matta2023digital,minerva2020digital,onaji2022digital,peng2023digital,semeraro2021digital,sharifi2024application,zhuang2021connotation}. 

The physical entity represented can include a machine, a ship, an entire city, or a human, such as a patient or a driver~\cite{bhatti2021towards,chen2024multiobjective,giering2021maritime,hu2022review,sharifi2024application}. 
For instance, a patient's condition or the status of an operation may be indicated~\cite{chen2024multiobjective}. Similarly, for a road user operating a vehicle, physical data, attention levels, and personality traits can be gathered through advanced algorithms and corresponding sensing technologies to accurately characterize the human driver~\cite{hu2022review}. A DT not only represents the state or 
characteristics of a physical entity but can also model it using 2D or 3D representations~\cite{bhatti2021towards,hu2022review,mohammed2022ontology,zhao2023digital}. 
For the user to interact with or observe the model, a human-machine interface (HMI) and data visualization capabilities may be necessary.

The DT quantifies measurable characteristics of a physical entity using distinct values that delineate various attributes of that entity. 
Typically, not all values are pertinent, and the pertinent ones may differ based on the application of the decision tree. These values 
enable the physical entity to be represented virtually, potentially on multiple occasions. Realistically and faithfully reflecting a 
physical entity in all its aspects is challenging, if not impossible; thus, it is represented virtually in an idealized, abstracted, 
and simplified form~\cite{giering2021maritime,jia2022simple,minerva2020digital,vanderhorn2021digital}.

A faithful virtual representation facilitates various features and applications, including effective monitoring, visualization, 
prediction, and simulations~\cite{giering2021maritime,park2019design,pires2020digital,wang2024pipeline}. This work distinguishes 
between simulation and simulations, despite the frequent conflation of the two terms.

Kritzinger et al.~\cite{kritzinger_digital_2018} utilize the \emph{virtual representation} to characterize a DM, representing the 
lowest tier among the three types of DT. Furthermore, in the context of maturity models, the ability to virtually represent the 
physical counterpart is often associated with the lowest maturity level of DTs~\cite{phua2022digital,zhao2023digital} (cf.~\cref{sec:background}. 
The \emph{virtual representation} is considered essential in all DM, DS, and DT. Only a DT and DS are capable of real-time replication 
due to automated data acquisition. A DM does not possess this feature and, consequently, lacks the capacity for real-time reflection~\cite{kritzinger_digital_2018,tao2018digital,vanderhorn2021digital}.

\subsubsection{Data Acquisition} 

A DT can autonomously gather and compile data about its physical counterpart and surroundings by utilizing sensors, in-situ sensors, 
embedded systems, and IoT devices that are linked to the digital entity and situated in the physical realm. Furthermore, integration of existing information systems, such as enterprise resource planning systems or third-party databases, may be required. These systems provide additional information, for example, on asset location (factory hall, manufacturing line, ...) or additional context information like current energy pricing, energy mix and $CO_2$ footprint as well as weather data. This approach allows for the input of substantial data volumes into the virtual entity, facilitating the documentation of 
incidents and alterations related to the physical entity~\cite{abdeen2023citizen,cimino2019review,cui2023physical,de2022digital,elmaraghy2022adaptive,
hribernik2021autonomous,levandovskiy2023experience,loaiza2023proposing,newrzella2022three,phua2022digital,schroeder2020methodology,vanderhorn2021digital,verna2024toward}.

Data acquisition establishes a link between the virtual and physical entity, preferably in real-time~\cite{aheleroff2021digital,loaiza2023proposing}. 
The data obtained may include details about the physical entity's state, behavior, or personality~\cite{brovkova2021implementation,hu2022review,matta2023digital}. 
Additionally, the sensors can collect and process ambient data that describes the physical entity's context and situation, indicating a degree of 
context awareness of the digital twin~\cite{hribernik2021autonomous,lim2020digital}. The data may encompass details about the location, objectives, 
jobs, or operators~\cite{hribernik2021autonomous,lim2020digital}. Real-time data acquisition facilitates various tasks, including prediction, analysis, 
and visualization~\cite{matta2023digital,phua2022digital}.

Automated data acquisition is identified as an essential component in the DT and DS feature model, as it signifies the core communication ability 
between virtual and physical entities. The most critical characteristic of a DS is its distinction from a DM. A DM, in contrast, can only receive 
data manually and lacks the capability to independently communicate with the physical entity. This feature is not included 
in the DM feature model, indicating that a DM does not incorporate this feature~\cite{kritzinger_digital_2018}.

\subsubsection{Synchronization}

\emph{Synchronization} ensures consistency and timeliness of virtual and physical entities by facilitating their appropriate 
updates~\cite{cimino2019review,jones2020characterising,minerva2020digital,wang2023digital}. Literature predominantly links synchronization to state changes in the physical entity as it is mirrored in the digital counterpart~\cite{matta2023digital,nwogu2022towards,peng2020digital}. 
The physical entity's condition is acknowledged, leading to an update of the digital entity~\cite{jones2020characterising}). This phenomenon 
is marked by varying twinning and update rates~\cite{jones2020characterising,matta2023digital,uhlenkamp2022digital}. The primary objective is 
to update the virtual entity in an event-driven and real-time manner immediately upon any change, ensuring that both entities remain equal and 
consistent at all times~\cite{jones2020characterising,li2022digital,madubuike2023scenarios,matta2023digital,peng2020digital,uhlenkamp2022digital,wang2023digital}. 
This enhances the dynamic nature of the DT's virtual representation, thereby improving its ability to depict reality in a timely and authentic 
manner~\cite{minerva2020digital,nwogu2022towards,uhlenkamp2022digital}. At a lower twinning rate, updates can be conducted at regular intervals, 
such as hourly or daily~\cite{uhlenkamp2022digital}. The required level of synchronization varies according to the application~\cite{minerva2020digital}. 
A lower update rate may be adequate in certain situations; however, in fields like medicine, near-instant synchronization is often essential to avoid 
the DT performing tasks based on outdated patient information~\cite{jones2020characterising,minerva2020digital,nwogu2022towards}.

Communication between virtual and physical entities facilitates their synchronization~\cite{liu20235m,peng2023digital}. \emph{Synchronization} 
is inherently linked to data acquisition and bi-directional communication. Given that a DS and a DT necessitate this form of communication, 
synchronization must be incorporated within them. A DM does not inherently engage with the physical entity, thus synchronization cannot be established~\cite{kritzinger_digital_2018}.

\subsubsection{Bi-directional Communication}

\emph{Bi-directional communication} is the most prevalent feature of DTs identified in the SLR, highlighting its significance and widespread 
acknowledgment (cf.~Fig.~\ref{fig:feature-count}). This feature is characterized by bi-directional automatic data exchange between the physical 
and virtual entities, allowing for mutual influence~\cite{ashraf2024digital,de2022digital,liu2023literature,nwogu2022towards,singh2024digital,vanderhorn2021digital,zhang2021energy}. 
Consequently, it relies on data acquisition. The virtual entity is generally updated by the physical entity, which is subsequently governed 
and influenced by the virtual model\cite{del2023digital,singh2024digital}. Literature indicates that this communication is characterized by 
being instant, automated, dynamic, seamless, and consistent~\cite{jeddoub2023digital,kamel2021digital,krishnamenon2021digital,minerva2020digital,ramu2022federated,singh2024digital}.

\emph{Data acquisition} represents one of the most critical aspects of a DT system. Bi-directional communication hereby serves as its defining 
characteristic, setting it apart from DMs, DSs, and other simulation frameworks~\cite{kritzinger_digital_2018}. Given that a DM is unable to independently communicate 
with the physical entity and a DS can only gather data, only a DT is capable of both receiving and transmitting data to 
and from the physical entity.

\subsubsection{Feedback~\&~Control}

\emph{Feedback and control} signify the communication process between the virtual entity of a DT and its physical counterpart~\cite{jones2020characterising}. 
This feature is distinct from bi-directional communication, though it suggests it, as numerous publications identify feedback or control 
mechanisms as a separate characteristic. Nonetheless, no instances were identified in which a DT includes feedback and control capabilities 
without facilitating bi-directional communication, thereby validating its classification in the GFM as a requisite child feature of 
bi-directional communication. The feedback and control capability, akin to bi-directional communication, is unique to DTs and is absent 
in DSs or DMs.

This feature involves digital-to-physical communication, encompassing control and feedback mechanisms related to the physical entity~\cite{hribernik2021autonomous,minerva2020digital,phua2022digital,popescu2022building,vanderhorn2021digital}. 
The DT is capable of transmitting data, commands, and updates to the physical entity, thereby exerting control and influence over 
it~\cite{de2022digital,minerva2020digital,onaji2022digital,vanderhorn2021digital,zeb2022industrial}. An AI-enabled Digital Twin can 
autonomously control the physical entity without user intervention~\cite{deren2021smart,lu2020digital,sharifi2024application}.
Decisions are typically made, or results produced within the DT, and necessary modifications to the physical entity are identified. 
These are accomplished via virtual processes, including data analytics and simulations. Subsequently, control mechanisms and actuators 
enable the virtual entity to update or adapt the physical entity in response to these changes, ideally in real time~\cite{han2023digital,jones2020characterising,
lu2020digital,luo2022digital,nwogu2022towards,vanderhorn2021digital}. 

A possible implementation of this communication process is feedback control, which represents a fundamental and automated version of a control 
loop~\cite{touhid2023building,hribernik2021autonomous,phua2022digital}. The feedback control mechanism can leverage data from the physical entity 
to adjust its parameter values accordingly~\cite{jones2020characterising,phua2022digital}. For example, when a specific reference value for car 
engine temperature is exceeded, simulations are conducted to assess various speeds and their effects. An optimal speed for reducing the temperature 
is determined, and the engine speed is subsequently adjusted~\cite{jones2020characterising}. Control of the physical entity can be achieved through 
both virtual processes within the DT and manual intervention by a user, such as remote control~\cite{de2022digital,nwogu2022towards}. Remote control 
is generally executed via a Human-Machine Interface (HMI), which signifies a unique characteristic within the GFM~\cite{aheleroff2021digital,lim2020digital,peng2020digital}. 

In conclusion, simulation capabilities or a specific level of cognition, such as data analytics, are necessary for generating decisions or 
results, thereby exerting an automatic influence on the physical entity~\cite{bhatti2021towards,han2023digital,jones2020characterising,van2023reference}. 
Additional features, such as AI or an HMI, may be necessary depending on the application.

\subsubsection{Interoperability}

\emph{Interoperability} involves proactive communication and exchange of data among Digital Twins, industrial and manufacturing applications, and 
systems like enterprise resource planning (ERP) systems~\cite{barricelli2019survey,de2022digital,hribernik2021autonomous,loaiza2023proposing,
park2019design,park2020operation,park2021vredi,wallner2023digital}. In the domain of manufacturing, this is also referred to as ``horizontal communication''~\cite{park2020operation,park2021vredi}. 
Interoperability enhances and supports context awareness of the DT~\cite{hribernik2021autonomous}. This is especially relevant for communication 
among DTs, as the various DTs serve as supplementary data sources for environmental information for one another~\cite{hribernik2021autonomous}.

Given that DTs and DSs facilitate automatic data exchange with the physical entity, this data flow can extend to other DTs, 
applications, or systems~\cite{kritzinger_digital_2018}. Thus, a DT and a DS can include this feature. The interoperability of a DS may be constrained 
by the inability to share data with other entities, which impedes effective communication~\cite{kritzinger_digital_2018}. This type of communication, 
supported by multiple publications~\cite{hribernik2021autonomous,pedersen2021living} signifies a potential advanced feature. Consequently, it is 
characterized as optional. A DM can not incorporate this feature due to the absence of automatic data exchange between its digital and physical 
entities~\cite{kritzinger_digital_2018}. Therefore, it follows that the DM is unable to interact with other applications and systems. Interoperability 
between a document management system and other systems can be attained through manual data flow by a user; however, this method alone is insufficient 
to fulfill the requirements of this feature.

\subsubsection{Internal Communication}
\label{ssec:int-comm}

DTs may comprise various models and sub-models, including geometric, behavioral, functional, and simulation models. Additionally, they encompass several 
components, such as data storage, sensors for data acquisition, external information systems and HMI~\cite{autiosalo2019feature,jia2022simple,kritzinger_digital_2018,zhuang2021connotation}. 
At an advanced level, a DT can incorporate various models corresponding to different life cycle phases of the physical entity, including design, operation, 
and disassembly models~\cite{schroeder2020methodology,zheng2022emergence}. Internal communication encompasses the connections, data transfer, and interactions 
among all these components~\cite{autiosalo2019feature,jia2022simple,luo2022digital,park2019design,semeraro2021digital}. For instance, it permits the utilization 
of data acquired from sensors or processed through virtual methods~\cite{de2022digital,autiosalo2019feature} introduced a feature that can be facilitated through 
a central data hub, enabling star-form communication that encompasses all data regarding the physical 
entity~\cite{giering2021maritime,pedersen2021living,tao2018digital}. Models can be constructed upon existing models, allowing for the access of stored data by 
other models and features. This also enables interaction between real-time and historical data~\cite{jia2022simple,pedersen2021living,tao2018digital}.

Autiosalo et al.~\cite{autiosalo2019feature} and others~\cite{brovkova2021implementation,pedersen2021living} identified the interaction of elements and data 
linkage as the most critical features of a DT. Additionally, it facilitates other derived features such as modularity and scalability, 
enhances data usability, and improves the authenticity of the DT~\cite{autiosalo2019feature,giering2021maritime,onaji2022digital,zhuang2021connotation}. 
Internal communication is essential for DTs and DSs, as sensor data must be accessible to other digital components, and the information derived from data 
analysis or simulations should be transferable to control systems or actuators~\cite{de2022digital,nwogu2022towards}. In contrast, a DM does not 
inherently gather or relay data to its physical counterpart~\cite{kritzinger_digital_2018}. The virtual entity acquires data manually, thereby eliminating 
the need for internal communication~\cite{kritzinger_digital_2018}. Consequently, it was conjectured that this feature is optional for a DM.

\subsubsection{Human-Machine Interface}
\label{sssec:hmi}

The HMI serves as the interactive communication channel connecting the DT with its stakeholders and users~\cite{autiosalo2019feature,
barricelli2019survey,de2022digital,schroeder2020methodology}. The main objective is to facilitate interaction between the user and the DT. 
Data can be made accessible to stakeholders and users through visualization and presentation via the HMI~\cite{aheleroff2021digital,de2022digital,pedersen2021living,schroeder2020methodology}. 
\cref{sssec:data-visuals} will provide a detailed explanation of data visualization. Conversely, users can engage with the DT through multiple 
methods via the HMI~\cite{autiosalo2019feature,loaiza2023proposing}. This involves modifying data visualization and conducting analyses and 
simulationsn~\cite{aheleroff2021digital,huang2022ethical,samak2023autodrive}. The HMI enables users to remotely control the physical entity 
via the DT and facilitates feedback and problem reporting~\cite{aheleroff2021digital,peng2023digital}).

Various types of HMIs exist, including graphical user interfaces and those utilizing virtual reality (VR) or augmented reality (AR)~\cite{autiosalo2019feature,schroeder2020methodology,wang2024pipeline}. 
A graphical user interface can be implemented via a basic web-based application, whereas more advanced VR or AR-enabled 
HMIs may require additional tools such as head-mounted displays~\cite{autiosalo2019feature}. The design and functionality of the HMI are contingent upon the 
users and application scenarios~\cite{autiosalo2019feature}.

\subsubsection{Extended Reality}

\emph{Extended reality} includes three advanced methods for HMI implementations: VR, AR, and mixed
reality~\cite{giering2021maritime,levandovskiy2023experience,schroeder2020methodology}. Consequently, the rationale applied to HMI can similarly 
be extended to this feature concerning its necessity and presence within the three feature models. Extended reality serves as a potential feature 
that may be integrated into all three types of Digital Twins. VR and AR are established yet sophisticated methodologies with a range of applicable 
scenarios~\cite{elmaraghy2022adaptive,popescu2022building}. 
Their provision of an interactive and immersive 3D experience can facilitate or augment new features and functionalities\cite{aheleroff2021digital,del2023digital,giering2021maritime}. 
VR and AR integrated within a DT can facilitate effective robot control through head-mounted displays~\cite{autiosalo2019feature}. In healthcare, 
physicians can utilize virtual reality to pre-experience and simulate a patient's surgery through a Digital Twin, effectively replicating elements 
such as blood flow and haptic feedback~\cite{chen2024multiobjective}.

Mixed reality seeks to integrate and link virtual components with the real world. This application may be utilized in testing scenarios for unmanned 
aircraft systems. The flight of a drone can be simulated in the digital twin, incorporating virtual obstacles that are not present in the drone's real 
flight environment. Utilizing mixed reality capabilities, the real drone navigates around obstacles in tandem with the virtual drone's movements in the 
virtual environment~\cite{zhao2023digital}.

\subsubsection{Data Visualization}
\label{sssec:data-visuals}

This feature allows the DT to visualize its data. This encompasses the visualization and presentation of real-time and historical data, data analytics 
outcomes, as well as 2D and 3D models, tables, and graphs~\cite{abdeen2023citizen,aheleroff2021digital,touhid2023building,ferreira2024digital,
matta2023digital,nwogu2022towards,pedersen2021living,schroeder2020methodology}. Consequently, data is rendered interactive and accessible to users, 
facilitating decision-making by providing real-time visibility into the condition of the physical entity~\cite{de2022digital,luo2022coupling,zhuang2021connotation}.

The virtual representation feature can exhibit a reciprocal dependency with data visualization. The HMI is involved and is typically necessary for 
rendering models of a physical entity visible to the user~\cite{touhid2023building}. Without an accurate virtual representation, the functionality 
and value of visualizations may be constrained~\cite{jeddoub2023digital}.

\emph{Data visualization}, as previously noted and corroborated by the literature~\cite{cimino2019review,de2022digital,pedersen2021living,schroeder2020methodology}, 
pertains to the HMI, which is essential for visualizing and presenting data to users. This supports the classification of data visualization as a 
fundamental aspect of the HMI. The primary function of a DM, DS, or DT is to virtually represent its physical entity~\cite{kritzinger_digital_2018}, 
and data visualization is often integrated as a fundamental component~\cite{wallner2023digital}. While an HMI can facilitate simulations or remote 
control of a physical entity, as outlined in~\cref{sssec:hmi}, data visualization is not essential for interactive communication between the DT and 
its ser~\cite{huang2022ethical,peng2023digital,samak2023autodrive}. Therefore, it is characterized as an optional feature.

%%MV
\subsubsection{Identification}

\emph{Identification} uniquely distinguishes both physical and virtual entities~\cite{autiosalo2019feature}. It consists of two types~\cite{autiosalo2019feature}. 
Physical identification universally identifies physical entities through methods like electronic product codes or radio 
frequencies, facilitating access to their digital counterparts~\cite{autiosalo2019feature,schroeder2020methodology}. Thus, it enables the linkage 
between the physical entity and its virtual counterpart~\cite{autiosalo2019feature,giering2021maritime,schroeder2020methodology}. The second type, 
digital identification,  uniquely identifies a virtual entity and facilitates access through a network, such as the Internet~\cite{autiosalo2019feature}. 
This capability can improve fleet coordination and facilitate the identification and integration of a DT into a hierarchical DT structure, such as 
that of a ship~\cite{giering2021maritime}. 

Identification, while infrequently addressed in the literature, is a prerequisite for DSs and DTs, facilitating communication between virtual and 
physical entities. Consequently, it is essential for data acquisition, synchronization, bi-directional communication, and possibly 
interoperability. The scarcity in the literature may stem from the perception that this feature is not regarded as distinct but rather as an 
enabling technology, a prerequisite, or an integral component of the communication feature. Due to the absence of automatic data flow, a DM cannot 
exhibit the specified communication features, leading to the conclusion that identification aspects are similarly excluded from its potential 
capabilities~\cite{kritzinger_digital_2018}.

\subsubsection{Data Processing}

\emph{Data processing} encompasses various functionalities that enhance the management and utilization of a physical entity's data~\cite{dihan2024digital,newrzella2022three}. 
Data is inherently heterogeneous, characterized by varying sources, dimensions, and timestamps~\cite{de2022digital}. Consequently, data must undergo 
processing, which includes functions such as filtering and aggregation~\cite{de2022digital}. To address high-dimensional data suitable data coding 
techniques should be incorporated~\cite{barricelli2019survey}. Homogenization facilitates interoperability and cost-efficient data management~\cite{loaiza2023proposing,wang2023digital}. 

Data fusion and ontologies are essential elements of data processing~\cite{barricelli2019survey,popescu2022building}. Data management and fusion techniques 
integrate diverse data types and consolidate various data sources and information sets related to the physical entity~\cite{barricelli2019survey,
deng2021systematic,newrzella2022three,uhlenkamp2022digital,van2023reference}. These data sources may encompass sensors, information systems, simulations, or 
analyses~\cite{uhlenkamp2022digital}. Consequently, redundancies may be reduced, leading to the extraction of more meaningful, consistent, and 
accurate information from the data, thereby enhancing its value for users~\cite{barricelli2019survey,newrzella2022three,uhlenkamp2022digital,vanderhorn2021digital}. 

Ontologies apply semantics to create a machine-readable vocabulary for data, facilitating its sharing with other systems or users~\cite{madubuike2023scenarios} 
facilitating cognitive reasoning and decision-making~\cite{zheng2022emergence}. Data processing can be employed to evaluate data quality from multiple sources, 
providing insights into the value added by sensor positions~\cite{pedersen2021living}. In cognitive DTs, data processing capabilities, including data fusion, 
can be improved through the application of artificial intelligence or machine learning, thereby increasing their intelligence~\cite{dalibor2022cross,veluvolu2024insight}.

Due to the intricate nature of data, and the necessity for all three types of DTs to manage and utilize the information and values of the physical 
entity for accurate representation, it is evident that data processing capabilities must be inherent in every digital model, shadow, and twin~\cite{de2022digital,minerva2020digital}.

\subsubsection{Data Storage}
\label{sssec:data-store}

Another characteristic of DTs is their capacity to memorize and store data. This includes multiple categories of physical entity data, such 
as time series, historical, current, static, dynamic, descriptive, and environmental data~\cite{barricelli2019survey,de2022digital,minerva2020digital,ramu2022federated}. 
A record of the changes made, and their outcomes concerning the physical entity, can be maintained~\cite{schroeder2020methodology}. Data storage 
can occur locally in record files or databases, or remotely via fog or cloud databases, facilitating quick and convenient access~\cite{autiosalo2019feature,pedersen2021living,schroeder2020methodology}. 
Furthermore, various data models may be employed for data storage, including SQL and NoSQL~\cite{autiosalo2019feature}. Providing the DT with 
access to historical data and protocols enables the execution of virtual processes, including simulations and diagnostics, based on this information~\cite{loaiza2023proposing}.

Storing extensive data is generally advantageous~\cite{minerva2020digital}. Due to the increasing volumes of data, efficient data storage 
is essential for retaining only relevant information~\cite{dalibor2022cross,schroeder2020methodology}. A potential technique for achieving this 
involves feature selection and extraction through machine learning algorithms~\cite{barricelli2019survey}. This technique minimizes data dimensionality 
by retaining only essential information, thereby reducing processing and storage expenses~\cite{barricelli2019survey}.

The literature indicates that data storage increases the value of a DT; however, its inclusion is not mandatory~\cite{minerva2020digital,schroeder2020methodology}. 
Additionally, there is no justification for claiming that data storage cannot take place in a DS and DM, making it optional for all three DT types.

\subsubsection{Simulation}
\label{sssec:sim}

\emph{Simulation} is a prevalent characteristic of DTs. Several studies~\cite{damjanovic2019open,park2019design,park2021vredi,uhlenkamp2022digital} identify 
it as a fundamental feature of Digital Twins. This aligns with the commonly accepted definition of DTs put forth by Glaessgen and Stargel in 2012~\cite{tao2018digital} 
and the prevalence of this characteristic in the literature  (cf.~Fig.~\ref{fig:feature-count}). Simulations may exhibit various characteristics, 
including static or dynamic forms, and the incorporation of probabilities~\cite{chen2022research,uhlenkamp2022digital,wang2024pipeline}. These tools are 
employed to replicate diverse configurations, scenarios, conditions, and alterations without affecting the physical entity~\cite{chen2024multiobjective,
ferreira2024digital,pedersen2021living,schroeder2020methodology}. What-if simulations enable precise execution of tests and experiments, allowing 
for the evaluation of outcomes to identify necessary modifications to the physical entity~\cite{barricelli2019survey,bauer2021urban,ferreira2024digital,lehtola2022digital}. 
Configurations of a production process can be adjusted without changing the actual process, the potential behaviors of individuals, and events within 
a city can be simulated to inform decision-making, and health-related tests and scenarios can be performed without impacting or harming a patient's 
body~\cite{chen2024multiobjective,deren2021smart,ferreira2024digital,schroeder2020methodology}. Simulations are commonly employed for predicting 
and estimating future states, failures, anomalies, behaviors, or performance of both physical entities and virtual commissioning, including validation 
and design optimization~\cite{autiosalo2019feature,damjanovic2019open,matta2023digital,minerva2020digital,nwogu2022towards,wang2023digital}.

Simulation is related to data analytics, however, it is a distinct concept. Predictive and prescriptive data analytics can utilize simulation 
capabilities, while data generated from simulations can also be used as input for data analytics tasks, and the relationship is reciprocal~\cite{autiosalo2019feature,nwogu2022towards,uhlenkamp2022digital}. 
Additionally, both features facilitate predictive and optimization capabilities, enhancing decision-making~\cite{barricelli2019survey,nwogu2022towards,pires2019digital}.

While simulation is frequently associated with DTs as a key characteristic, it is not an essential component for all DTs, DSs, or 
DMs~\cite{park2020operation,uhlenkamp2022digital}. A decision model and decision support may incorporate simulation models~\cite{kritzinger_digital_2018}. 
This is justified as they can virtually represent the physical entity or analyze its data, which does not inherently necessitate the 
application of simulations~\cite{kritzinger_digital_2018}. DTs must possess the ability to autonomously update or influence their physical 
counterparts~\cite{minerva2020digital,nwogu2022towards}. This requires the generation of decisions or outcomes, which is facilitated by 
virtual processes such as simulations or data analytics~\cite{han2023digital,jones2020characterising,lu2020digital,luo2022coupling,nwogu2022towards,vanderhorn2021digital}. 
Thus, as illustrated by the or-constraint of the DT feature model in Fig.~\ref{fig:dt}, each DT must incorporate simulations or a specific level 
of cognition, indicating the necessity of data analytics.

\subsubsection{Cognition}

Cognition is an abstract parent feature, i.e., not implementable independently, that includes 
three sub-features: AI, data analytics, and machine learning, with the latter categorized as a sub-feature of data 
analytics. Uhlenkamp et al.~\cite{uhlenkamp2022digital} assert that data analytics represents a manifestation of 
cognition and intelligence related to digital technologies. This corresponds with the feature model of a DT-based 
predictive maintenance system proposed by van Dinter et al.~\cite{van2023reference}, in which the analysis is 
classified as a sub-feature of cognition. The differentiation between AI and ML is established in the work 
of Autiosalo et al.~\cite{autiosalo2019feature}) as they provide varying capabilities, defining AI as related to autonomous decision-making, rendering the DT self-active. 
Conversely, machine learning is employed to process and analyze data for users or decision-making, differentiating it from artificial intelligence and situating it within the realm of data analytics. 
Data analytics encompasses descriptive, diagnostic, predictive, and prescriptive capabilities, with machine learning 
functioning as a sub-feature that supports and enhances these aspects~\cite{ramu2022federated,uhlenkamp2022digital}.

\emph{Cognition} is not a prerequisite for the existence of a DT. Once a specific level of cognition is attained, data analytics becomes 
essential, as descriptive analytics signifies a fundamental level of cognition~\cite{uhlenkamp2022digital}. Multiple studies~\cite{elmaraghy2022adaptive,mingorance2023evolution,popescu2022building} 
indicate that data analytics is generally employed in a Digital Twin prior to the application of machine learning or artificial intelligence.
Data analytics is an essential component of cognition, while machine learning and artificial intelligence are supplementary.

\subsubsection{Data Analytics}

This feature includes descriptive, diagnostic, predictive, and prescriptive analytics~\cite{onaji2022digital,uhlenkamp2022digital,wang2023digital}. 
This may allow the DT to offer insights into historical events, expected future occurrences, underlying causes, or recommendations concerning the 
physical entity~\cite{uhlenkamp2022digital}. Data analytics has multiple applications. Data from past or real-time behavior of a physical 
entity can be sourced and analyzed to monitor its current state and condition~\cite{bauer2021urban,matta2023digital,pires2020digital}. Analyzing 
and diagnosing errors associated with digital or physical entities allows for the identification of necessary changes or modifications~\cite{pedersen2021living}. 
Predictive and forward-looking analysis facilitates the detection or estimation of anomalies and failures within a physical entity prior to 
their occurrence, thereby enabling prevention~\cite{barricelli2019survey,hassani2022impactful,pires2020digital}. In practical applications, 
it is possible to predict a patient's health status, anticipate future events in a manufacturing process, and estimate the timing of the next 
required maintenance for a machine~\cite{barricelli2019survey,chen2024multiobjective,damjanovic2019open,ferreira2024digital}. Descriptive or 
prescriptive analysis results can subsequently inform prescriptive analytics, which entails decision-making or offering recommendations and 
actions~\cite{barricelli2019survey}. %Machine learning and artificial intelligence can further enhance data analytics, as discussed in the 
%subsequent sections~\cite{giering2021maritime}.

Data analytics, as discussed in~\cref{sssec:sim}, is closely related to simulation, yet fundamentally distinct. Both features can 
function as interdependent data sources, enabling predictions, optimizations, and informed decision-making~\cite{autiosalo2019feature,
barricelli2019survey,nwogu2022towards,pires2019digital,uhlenkamp2022digital}. Data analytics can enhance optimization and support decision-making 
through sensitivity or correlation analyses~\cite{autiosalo2019feature,giering2021maritime,hassani2022impactful,mohammed2022ontology,zhuang2021connotation} .

Data analytics serves as an optional feature in all DM, DS, and DT frameworks; however, it becomes a requisite upon reaching a specified 
level of cognition, as detailed in the preceding section~\cite{uhlenkamp2022digital}. Kritzinger et al.~\cite{kritzinger_digital_2018} describe 
that a DM and DS may incorporate mathematical models, though this is not a requirement. This justification arises from their potential use in 
accurately representing physical entities, which does not inherently require data analytics~\cite{kritzinger_digital_2018,matta2023digital,minerva2020digital}. 
Furthermore, certain literature suggests that a DT can represent a maturity level that lacks data analysis capabilities~\cite{elmaraghy2022adaptive,matta2023digital,minerva2020digital}. 
The justifications for the existence and necessity of simulations in a digital twin also apply to the chosen representation of data analytics 
within the DT feature model.

\subsubsection{Machine Learning}

According to Autiosalo et al.~\cite{autiosalo2019feature}, \emph{machine learning} is often used to process and analyze data for the user 
or AI, supporting decision-making and setting it apart from AI. While the authors of this work share this perspective, it is still important 
to note that several publications do not differentiate between ML and AI to this degree~\cite{barricelli2019survey,popescu2022building}. As a 
child feature of data analytics, it supports various aspects of data analytics, such as prediction and optimization~\cite{abdeen2023citizen,giering2021maritime,pires2019digital,ramu2022federated}. 
By employing machine learning through supervised and unsupervised learning algorithms, including deep learning, the DT can continually 
self-learn and advance by processing data from the physical entity~\cite{aheleroff2021digital,barricelli2019survey,damjanovic2019open}. For 
instance, neural networks or support vector machines can be used to learn from training data, identify patterns, and subsequently conduct 
predictions~\cite{levandovskiy2023experience,wang2024pipeline}. A practical use case of machine learning can be the application of a DT 
in a welding process~\cite{lu2024temperature}. Once real-time data on the welding temperature is received, support vector regression can 
be employed to accurately predict the temperature in the core area and to activate an alarm if the temperature surpasses a predefined 
threshold~\cite{lu2024temperature}. Another use case involves feature selection and extraction to minimize data storage and processing 
resources, as previously outlined in~\cref{sssec:data-store}~\cite{barricelli2019survey}.

Like data analytics, \emph{machine learning} can be a potential feature of a DM, DS, and DT, as a DM can already encompass mathematical 
models. Additionally, cognition is independent of the level of data integration, which distinguishes the three types of DTs~\cite{kritzinger_digital_2018}. 
Furthermore, machine learning is an optional child feature of data analytics. ML closely relates to data analytics and supports it, but is 
usually implemented at a more advanced maturity stage than data analytics, according to several researchers~\cite{autiosalo2019feature,
elmaraghy2022adaptive,giering2021maritime,mingorance2023evolution,popescu2022building}. Consequently, machine learning necessitates data 
analytics and cannot exist in a DM, DS, or DT without it.

%%mv_STOP
\subsubsection{Artificial Intelligence}

This work's understanding of AI corresponds with the concept articulated by Autiosalo et al.~\cite{autiosalo2019feature}, which posits that 
AI pertains to autonomous decision-making, thereby rendering the DT self-active. Therefore, AI can be differentiated from ML~\cite{autiosalo2019feature}. 
AI enables a DT to make autonomous decisions, including reacting to events, controlling physical entities, and issuing intelligent early 
warnings~\cite{autiosalo2019feature,hribernik2021autonomous,loaiza2023proposing,mingorance2023evolution,phua2022digital,sharifi2024application,zhao2023digital}. 
Autonomy is partially influenced and strengthened by advanced cognition; therefore, high-level cognition was also considered an application of AI~\cite{zheng2022emergence}. 
This enables a decision tree to perform processes analogous to human cognition, such as reasoning, learning, problem-solving, decision-making, 
and the recognition of anomalous behavior~\cite{barricelli2019survey,elmaraghy2022adaptive,zheng2022emergence}. 

AI and autonomous decision-making enable a digital system to operate independently. Another important functionality is self-evolution and 
co-evolution with the physical entity~\cite{autiosalo2019feature,barricelli2019survey,elmaraghy2022adaptive,mihai2022digital}. Dynamic self-evolution 
encompasses autonomous monitoring of the physical entity and its environment, self-assessment and diagnosis, self-optimization, and, crucially, 
self-adaptation~\cite{lv2023bio,mihai2022digital,wang2024framework}. A DT autonomously and dynamically modifies, updates, and enhances its behavior 
and models in response to changes in the physical entity and its environment. This process facilitates model growth and ensures the virtual entity 
remains accurately aligned with its physical counterpart throughout its life cycle~\cite{elmaraghy2022adaptive,lv2023bio,guo2022application,
hribernik2021autonomous,matta2023digital,peng2023digital,tao2018digital,wang2024framework,zheng2022emergence}. This advanced cognitive ability 
transcends basic synchronization and is pertinent, for instance, when the physical entity is downgraded or an additional element is introduced~\cite{elmaraghy2022adaptive,matta2023digital}. 
Dynamic self-evolution is considered a fundamental attribute of DTs~\cite{tao2018digital,wang2024framework}. According to Hribernik et al.~\cite{hribernik2021autonomous}), 
self-adaption is associated with autonomy and necessitates decision-making techniques.

The AI-related aspects discussed, such as decision-making, autonomy, adaptation, cognition, and dynamic evolution, exhibit significant overlap 
and interconnection, and can be enhanced through simulations and data analytics~\cite{autiosalo2019feature,barricelli2019survey,hribernik2021autonomous,uhlenkamp2022digital,zheng2022emergence}.

In the literature, artificial intelligence is frequently characterized as a sophisticated attribute indicative of the peak maturity of a DT, 
rendering it an optional component~\cite{elmaraghy2022adaptive,mingorance2023evolution,phua2022digital,zhao2023digital,zheng2022emergence}. The 
reasons provided for the existence and necessity of data analytics and machine learning in feature models also substantiate the inclusion of AI 
as an optional feature in data management, data science, and data technology. The degree of functionality can vary considerably among the various 
types of DTs. A DM has restricted applications of AI due to the absence of automatic linkage to its physical entity.

\subsubsection{Computation}

Computation is a critical aspect that is infrequently addressed in the literature. This may be attributed to its position at a low 
hierarchical level, which many researchers might overlook. Nonetheless, it is included in the feature models for the sake of completeness. 
Computation produces data through the resolution of mathematical problems. It may occur locally, reducing the delay in data exchange between 
virtual and physical entities, or globally, providing extensive data processing resources~\cite{autiosalo2019feature}.

Computation is deemed essential for all decision makers, data scientists, and data technicians due to its fundamental role in addressing 
mathematical tasks and its low hierarchical level. It serves as a fundamental component for additional features, including data processing, 
which is essential across all three types of Digital Twins~\cite{autiosalo2019feature}.

\subsubsection{Security}

DTs interact with the data of their physical counterparts, exhibit heterogeneity, and frequently function in cyberspace, making them especially 
susceptible to cybersecurity challenges~\cite{autiosalo2019feature,de2022digital,giering2021maritime}. Potential security risks encompass unauthorized 
access or manipulation of data, theft, and denial-of-service attacks~\cite{autiosalo2019feature,de2022digital}. Cybersecurity protection measures can 
be employed to reduce the risk associated with these vulnerabilities~\cite{autiosalo2019feature,de2022digital,popescu2022building}. Access control is 
the most frequently cited measure in the reviewed literature~\cite{de2022digital,giering2021maritime,schroeder2020methodology}. Access control can include 
role-based access, restricted access to particular data or functionalities, access lists, or an authentication process to ensure data confidentiality~\cite{de2022digital,giering2021maritime,schroeder2020methodology}. 
Additional measures encompass data encryption and implementation of systems designed to detect suspicious activities or prevent data leakage~\cite{de2022digital,giering2021maritime}.

Literature and definitions indicate that security is a relevant, albeit not essential, characteristic of DTs~\cite{de2022digital,giering2021maritime,kritzinger_digital_2018,schroeder2020methodology}. 
Popescu et al.~\cite{popescu2022building} further substantiate this, characterizing cybersecurity as an aspect that digital technologies should include, 
though it is not strictly essential. Security measures operate independently of the data exchange between physical and digital entities, suggesting 
that this feature may be present in a DM, DS, and DT~\cite{kritzinger_digital_2018}.

\subsubsection{Modularity}

Modularity facilitates integration, addition, and exchange of models within a DT~\cite{semeraro2021digital}. The approach segments 
the DT into interconnected, reusable functional units, thereby enhancing flexibility and streamlining its design~\cite{loaiza2023proposing,semeraro2021digital}. 
This enhances its comprehensibility~\cite{loaiza2023proposing}. Modularity facilitates the reuse and rearrangement of units, thereby enhancing 
the adoption of DTs across diverse applications~\cite{loaiza2023proposing,mihai2022digital}.

The definitions provided by Kritzinger et al.~\cite{kritzinger_digital_2018} and the existing literature do not offer clear guidance on the 
classification of this feature. It can be inferred from the literature that modularity is not essential in a DT. Furthermore, there is no justification 
for claiming that this feature cannot also manifest in a DS.

\subsubsection{Scalability/Composability}

\emph{Scalability} enables a DT to provide information regarding a physical entity across multiple scales~\cite{caprari2022digital,semeraro2021digital}. 
Consequently, the physical entity may be represented and analyzed across various levels~\cite{matta2023digital}. The levels may denote various system tiers, 
including units, systems, and systems of systems, or different levels within a manufacturing plant, such as individual machines or the entire facility~\cite{matta2023digital,uhlenkamp2022digital}. 
For example, when the physical entity of a DT is a manufacturing plant, and the entire plant is the focus, the DT offers information including the 
maintenance schedule for the plant~\cite{jia2022simple}. However, if the object of interest is a specific production line at the plant, the DT adaptively 
modifies the information it provides~\cite{jia2022simple}. The DT can provide a comprehensive assessment of a machine or component's health status~\cite{jia2022simple}.

\emph{Composability} enables a DT to integrate various physical entities into a singular, comprehensive physical entity, which is overseen or 
managed via a unified virtual representation. In a manufacturing plant, various machines comprise a production line, while multiple production 
lines make up the entire facility. Each machine exists as a physical entity with an associated digital representation. All machines across 
production lines can be integrated into a unified virtual representation of the entire manufacturing plant, encompassing the virtual models of 
both the production lines and the machines. The entire plant can be monitored and analyzed through a comprehensive virtual representation that 
integrates all its distinct components. A complex system can be effectively represented, including its subsystems and relevant information~\cite{guo2022application,
jia2022simple,minerva2020digital,park2021vredi}.

Scalability and composability represent distinct dimensions of the same characteristic. To integrate diverse digital entities representing 
various physical entities, internal communication is essential to facilitate interaction among their representations~\cite{minerva2020digital}.
Both represent advanced features that enhance the value of a Digital Twin; however, they are not deemed essential, as indicated by certain 
researchers~\cite{matta2023digital,minerva2020digital}. It is assumed that a virtual representation of a complex physical entity, including 
multiple subsystems, can be created and utilized even in the absence of automatic data exchange regarding the physical entity~\cite{minerva2020digital}. 
Consequently, it was determined that this represents a potential feature in a DM, DS, and DT.

\subsection{Comparison of Covered Application Domains Related to the Derived Features}
\label{ssec:domains}

\begin{figure*}[bth]
\centering
    \includegraphics[width=.8\textwidth]{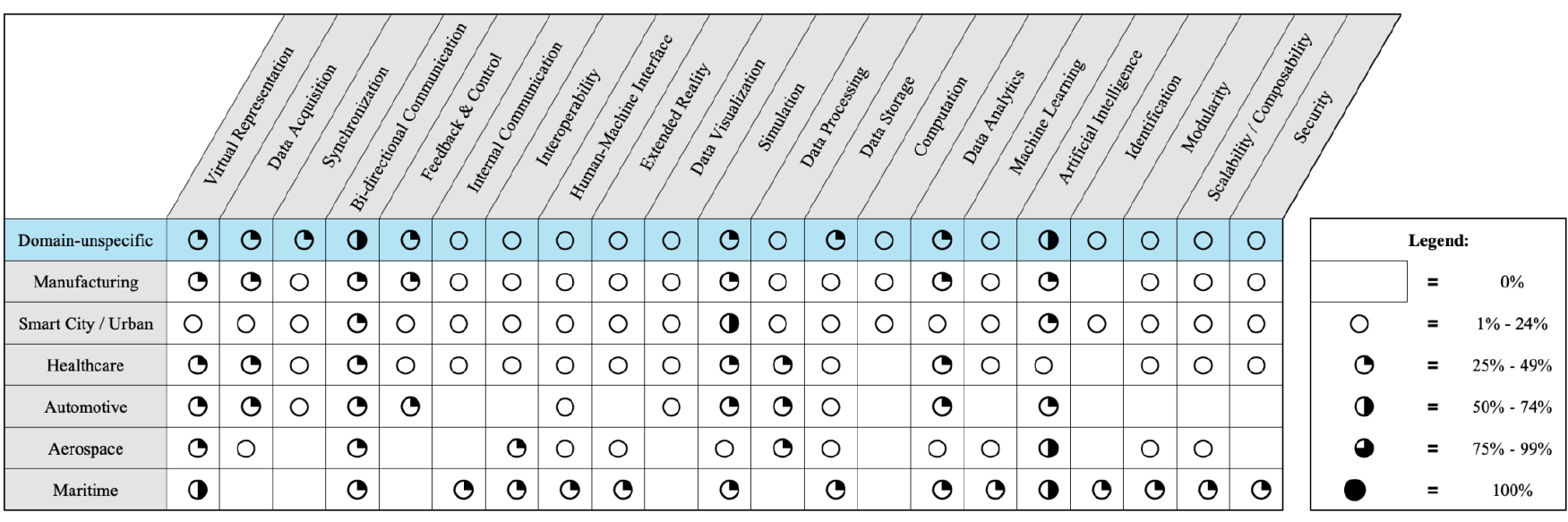}
    \caption{Occurrence of derived features across application domains in relation to publication (counts in percentage).}
    \label{fig:feature-domains}
\end{figure*}

The developed GFMs encompass various domains. Consequently, they cover the primary application domains of DTs: manufacturing, healthcare, 
smart cities, automotive, aerospace, and maritime. Additionally, we considered domain-agnostic features because of their universal applicability. Furthermore, we developed a matrix (cf. Fig.~\ref{fig:feature-domains}) to enhance the meaningfulness and transparency of the feature models, illustrating the distinctions among the application 
domains concerning the DT features derived from the SLR. The presence of each feature in each 
application domain can be inferred, facilitating a comparison of results across these domains.

% Fig.~\ref{fig:feature-domains} illustrates the frequency of each derived feature concerning the number of publications that reference DT 
% features within a specific application domain. 
Fig.~\ref{fig:pubs-domain} provides the necessary data to ascertain the number of publications 
within each application domain. %The representation of feature 
% appearances in application domains is illustrated through various types of circles. The frequency of features in the corresponding publications 
% varies according to the abundance of the circles, as indicated by the legend in Fig.~\ref{fig:feature-domains}. 
Virtual representation was 
identified in 25\% to 49\% of all manufacturing publications as a Digital Twin feature, whereas it appeared in only 1\% to 24\% of all smart city publications, 
according to the data presented in Fig.~\ref{fig:pubs-dom-feature}. Figs.~\ref{fig:pubs-dom-feature} and~\ref{fig:feature-count} illustrate the 
variability in the frequency of features and application domains. The significance of Fig.~\ref{fig:pubs-dom-feature} is confined to the comparative 
analysis of application domains in relation to their features. Additionally, the infrequent presence of computation, identification, modularity, 
and security features, coupled with the limited publications in the aerospace and maritime sectors, as illustrated in Figs~\ref{fig:pubs-dom-feature} 
and~\ref{fig:feature-count}, necessitates a cautious interpretation of these features and domains. As a result, they are afforded diminished attention 
in the subsequent assessment.

Fig.~\ref{fig:pubs-dom-feature} illustrates that the majority of features are depicted by either hollow or quarter-filled circles throughout 
the domains. Consequently, most features manifest infrequently or to a moderate extent within the application domains. In instances where a 
circle is not visible, the feature is omitted from the relevant domain. Fig.~\ref{fig:feature-domains} shows only six cases with a half-filled 
circle, and there 
are no instances of three-quarter-filled or completely filled circles. This indicates that no feature is consistently prominent or essential 
across the examined domains, likely owing to the significant diversity in the application scenarios of digital technologies.

The comparison of application domains reveals that the differences among them are predominantly minimal. Bi-directional communication, 
virtual representation, simulation, and AI are frequently referenced across various domains, highlighting their significance for DTs. Conversely, 
aspects such as data visualization, scalability, internal communication, and HMI are referenced infrequently or omitted entirely across most domains. 
Consequently, most features manifest consistently across various application domains, irrespective of their frequency of citation. While the number 
of publications in the aerospace and maritime domains is limited, they sufficiently correspond with other domains in terms of feature occurrence. 
Nonetheless, certain exceptions are present. Smart city publications reference virtual representation, data acquisition and analytics, and feedback 
and control less frequently than other domains. They discuss simulation more than any other domain as a feature. Smart city Digital Twin applications 
primarily emphasize simulations, including the modeling of scenarios and the anticipated behavior of individuals and events within urban environments~\cite{deren2021smart,lehtola2022digital}. 
AI and autonomous decision-making appear to be less linked to healthcare compared to other domains. The automotive domain exhibits the lowest 
diversity of features relative to other domains. Consequently, it can be inferred that the automotive sector presents limited diversity in use 
cases. It demonstrates minimal deviations from other domains.

Upon examination of the domain-unspecific features, it is evident that they largely maintain consistency with the covered application domains. 
Moreover, while the quantity of domain-unspecific publications is similar to that of smart cities and approximately half that of manufacturing 
publications, it represents the sole domain that includes all 21 derived DT features. These observations underscore two essential 
points. Firstly, none of the features is limited to a specific domain, and secondly, all features should be applicable to 
other application domains of DTs not addressed in this thesis.

In conclusion, the differences among the application domains discussed are primarily minimal, and the derived features, as well as the developed 
feature models, should exhibit a significant level of universality. It is essential to recognize that the results are significantly affected by 
the number of features cited in each publication and the number of publications within each domain.

\section{Applying the Feature model}
\label{sec:application}

In this section, we apply the developed feature models (cf.~\cref{sec:models}) to three use cases across 
various application domains, illustrating their flexible applicability, capacity to differentiate between 
digital models, shadows, and twins, and the interconnectivity and interdependency of the derived features. 
The use cases are presented in a simplified manner, emphasizing the derived DT features. While, for the sake 
of space, the use cases are not discussed in their entirety, as described by the authors, they cover the parts 
vital for our feature models confirming their applicability for diverse domains and application cases 
(Lu et al.~\cite{lu2024temperature}; Wang~\cite{wang2024digital}; Zhang et al.~\cite{zhang2022building}).
The use cases were chosen based on their diversity in application, goal, and domain.

\subsection{Fire Identification in Buildings}

Our first use case of DTs pertains to the domain of smart cities and urban applications. The objective is 
to detect and represent a building fire in real time. As part of this, a decision tree is employed to gather, analyze, 
and represent fire data effectively. This use case is essential due to the life-threatening and costly nature 
of fires in buildings and urban areas, providing firefighters with critical 
information and facilitating decision-making before entering hazardous areas~\cite{zhang2022building}.

Zhang et al.~\cite{zhang2022building} identified five core elements in fire identification through a decision tree: a fire within a room as the physical entity, IoT sensors, a cloud server, an AI engine, and an HMI. Effective fire identification requires the presence and interaction of several derived features. During a fire incident, data is first gathered by IoT sensors, including temperature sensors located in the room. This data is then transmitted wirelessly and in real-time to the cloud server of the DT, which assumes features such as data acquisition, internal communication, synchronization, and identification. The data is stored and processed on a cloud server, utilizing machine learning algorithms that necessitate a specific level of computation. Multiple simulations of virtual fire scenarios must be performed to develop a training dataset for the neural network of the DT. This machine learning approach and data analytics enable the identification and prediction of fire and its growth with high accuracy. This process may incorporate artificial intelligence and autonomous decision-making. 

Following this, the fire's location within the room, the heat release rate, and additional pertinent information can be modeled virtually and transmitted through internal communication processes to the HMI, where the data is visualized for firefighters as graphs or coordinate systems with a time lag of under one second. Moreover, the scalability of fire identification can range from a single room to a comprehensive system encompassing an entire building, while incorporating additional features such as virtual reality~\cite{zhang2022building}.

\begin{figure}[bth]
\centering
    \includegraphics[width=\columnwidth]{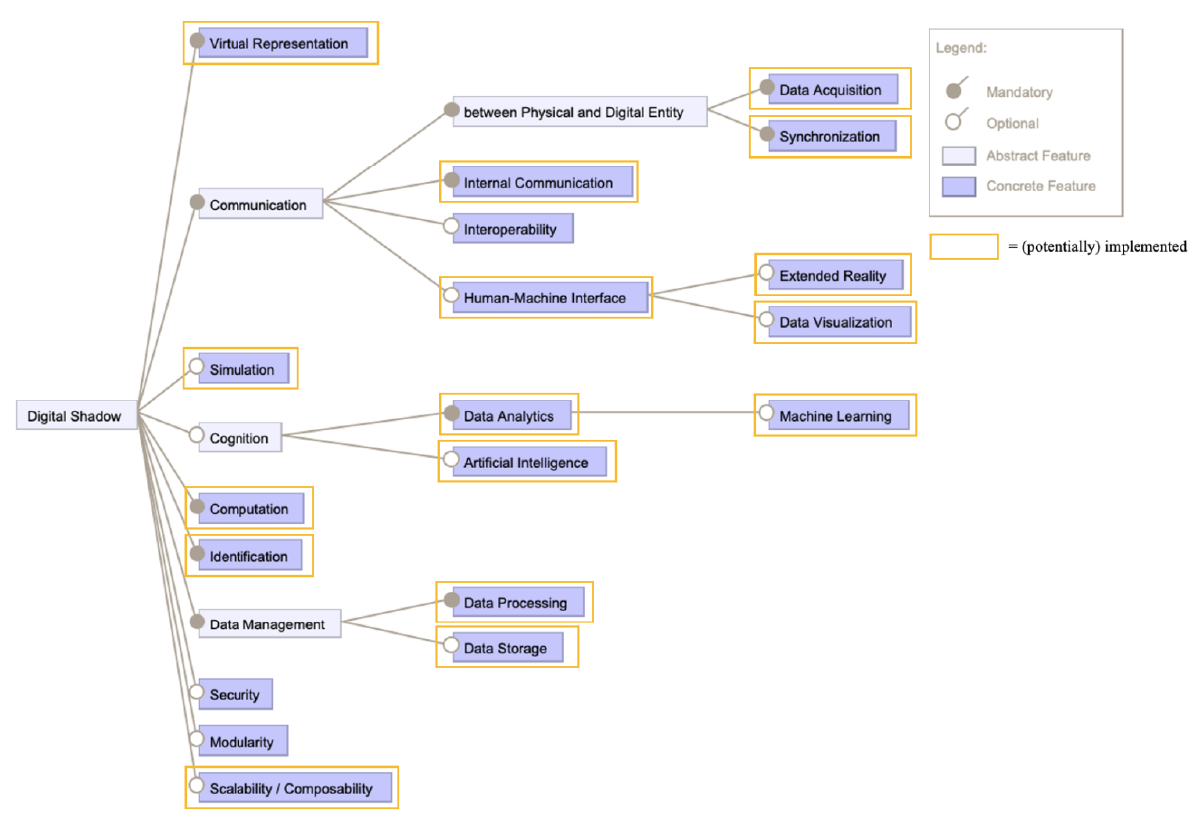}
    \caption{Feature Model of a DS for Fire Identification in Buildings: 16 (Potentially) Implemented Features; Created with FeatureIDE (Eclipse).}
    \label{fig:fire}
\end{figure}

The use case illustrates that detecting fires via a decision tree is not a simple task that can be achieved through a single feature of the decision tree. Several features of the DT feature model must or can play a role in this application (cf.~Fig.~\ref{fig:fire}). This use case demonstrates the interdependence of features within the feature model, highlighting their cumulative nature The simulation feature enhances machine learning, whereas the HMI is crucial for the real-time communication of analyzed data to firefighters. Bi-directional 
communication and control over the physical entity -- two essential features of DTs as specified in the developed feature model -- are absent in the fire identification process presented by~\cite{zhang2022building}. Fig.~\ref{fig:fire} illustrates that the use case corresponds with the characteristics of a digital shadow according to the digital shadow 
feature model (cf.~Fig.~\ref{fig:ds}). Consequently, while characterized as a DT application~\cite{zhang2022building}, the findings of this thesis suggest that it more appropriately represents a use case of a DS. This illustrates the interchangeable application of the terms digital model, digital shadow, and DT in the literature~\cite{kritzinger_digital_2018}. 
A decision tree would be essential if the use case includes control mechanisms, such as activating devices like sprinklers in the relevant room.

\subsection{Collision~\&~Conflict Warning for Vehicles}

The second use case involves a collision and conflict warning system for vehicles, specifically within the automotive domain. The increasing prevalence of autonomous and intelligent vehicles has led to a 
heightened demand for effective safety systems. Similar to the previous use case, a decision tree is utilized to create a collision and conflict warning system for intelligent vehicles, which can analyze and predict traffic scenarios while offering suitable recommendations to avert collisions~\cite{wang2024digital}.

Wang et al.~\cite{wang2024digital} state that data from actual vehicles and their surroundings are continuously gathered via sensors, devices, and GPS to develop a virtual model of traffic to generate collision and conflict alerts. This incorporates a degree of synchronization and identification capabilities. The collected 
data is analyzed to extract insights regarding driver behaviors, which are retained through data storage. Data classification and prediction can be performed using computation, data analytics, and machine learning algorithms, including k-means, linear regression, and decision trees. Consequently, it is possible to predict potential driver behavior and actions prior to their occurrence, classify and identify dangerous drivers, and ascertain and visualize hazardous zones. Nonetheless, the findings of this thesis indicate that a form of HMI is necessary for data visualization. Integrating simulation capabilities enables various scenarios and experiments, facilitating 
deeper insights and results. AI can subsequently be utilized in the decision-making process of the DT to activate alarms based on analysis and predictions. The DT should possess the ability to self-learn and self-optimize its parameters in response to incoming data, exemplifying an additional implementation of AI. The hazardous zones and preventive measures to avoid collisions are conveyed to physical entities, such as vehicles and pedestrians. Consequently, affected vehicles may be required to undertake specific measures to prevent a forecasted situation. This process exemplifies bi-directional communication and feedback to physical entities. Successful implementation of a collision and conflict warning system in vehicles requires the DT system to communicate with diverse real-world systems and objects, including other intelligent vehicles, infrastructure devices, networks, and smartphones, to collect relevant 
data and provide recommendations to all involved parties. The DT must facilitate internal communication to enable data exchange among sensors, data analysis, data storage, and data visualization. Consequently, interoperability and internal communication are essential components of the Digital Twin for executing this use case~\cite{wang2024digital}.

\begin{figure}[bth]
\centering
    \includegraphics[width=\columnwidth]{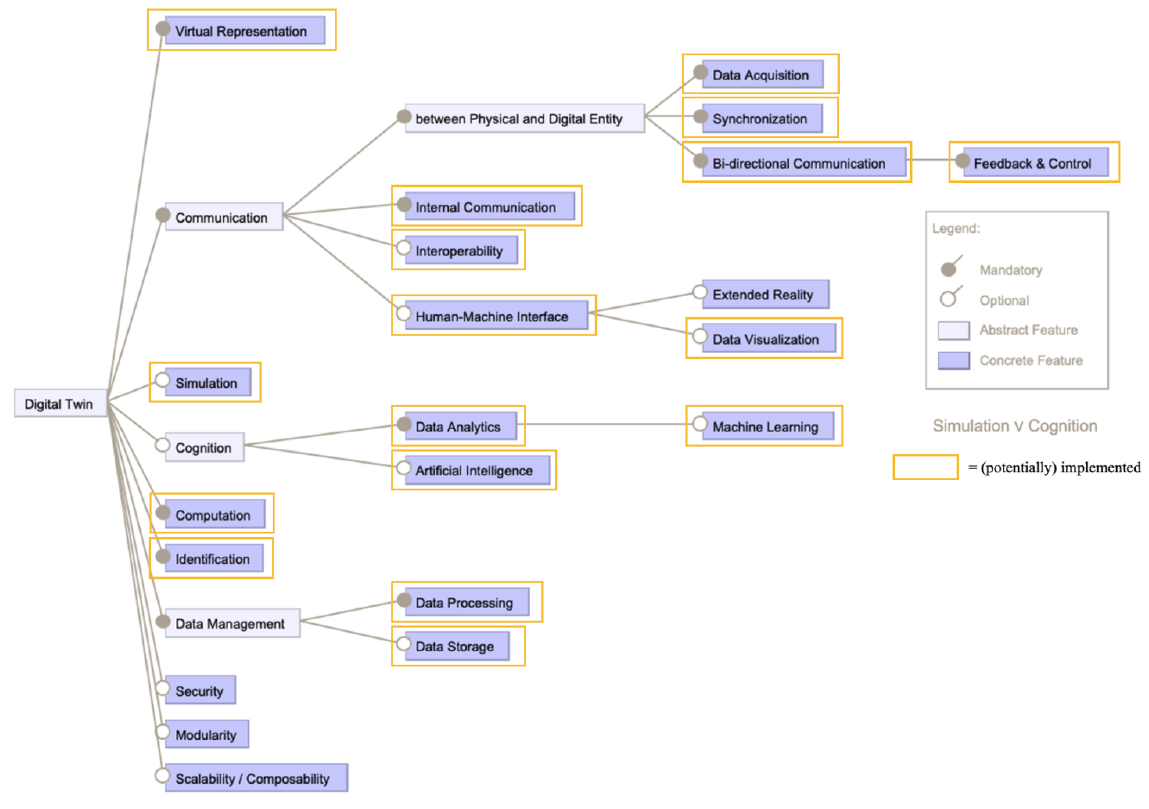}
    \caption{Feature Model of a DT for Collision~\&~Conflict Warning for Vehicles: 17 (Potentially) Implemented Features; Created with FeatureIDE (Eclipse).}
    \label{fig:collision}
\end{figure}

Collision warning, akin to the prior use case, entails a complex process that incorporates nearly all interdependent features of the DT feature model, as illustrated in Fig~\ref{fig:collision}. A collision warning system necessitates feedback to the physical entity, thereby incorporating all essential features of DT as outlined in the proposed feature 
model within this work (cf.~\cref{sec:models}). This use case can only be executed through a DT, excluding a digital shadow or model.

\subsection{Temperature Monitoring in Welding Processes}

Monitoring processes or assets represents a common application of DTs across various domains. It can pursue multiple objectives and, consequently, include various characteristics~\cite{touhid2023building,han2023digital,lu2024temperature,wang2024pipeline}. This third use case pertains to temperature monitoring in friction stir welding processes in the manufacturing sector. 
The quality of a weld is directly associated with the welding temperature, making this factor essential for welding tasks~\cite{lu2024temperature}.

Lu et al.~\cite{lu2024temperature} introduced a framework for monitoring welding process temperature, comprising five components: a physical entity, a digital entity, data connections, a digital twin-based friction stir welding system, and services. The DT-based welding system was developed from an engine that enables flexible augmentation of units and 
functions, thereby incorporating a level of modularity. In the welding process, data including surface temperature and parameters are collected via sensors from the welding equipment and transmitted to the DT-based welding system in real-time. This data is utilized to create a virtual replica, specifically a 3D model, of the friction stir welding tool and 
process. Motion simulation facilitates synchronization and alignment with the physical entity and its environment. A simulation of the temperature field related to the weldment can also be performed. Surface temperature data obtained from sensors can be utilized with Support Vector Regression, a machine learning algorithm, to predict the core temperature 
related to the welding process. An alarm activates when the forecast suggests a temperature surpassing a specified threshold. Given that the 3D model and the machine learning algorithm incorporate multiple models, such as geometric and physical models, and that various communication channels between the components are established, internal communication 
capabilities are crucial in this context. The communication channel between physical and digital entities enables the digital entity to collect data from the physical entity and return analysis or simulation results. The DT-based system facilitates bi-directional communication and feedback. The acquired data and prediction results concerning 
core temperature can be visualized in an HMI for users. The use case implies the inclusion of identification, computation, data analytics, and data processing through the automatic communication between digital and physical entities and the application of a machine learning algorithm, despite not being explicitly stated~\cite{lu2024temperature}.

\begin{figure}[bth]
\centering
    \includegraphics[width=\columnwidth]{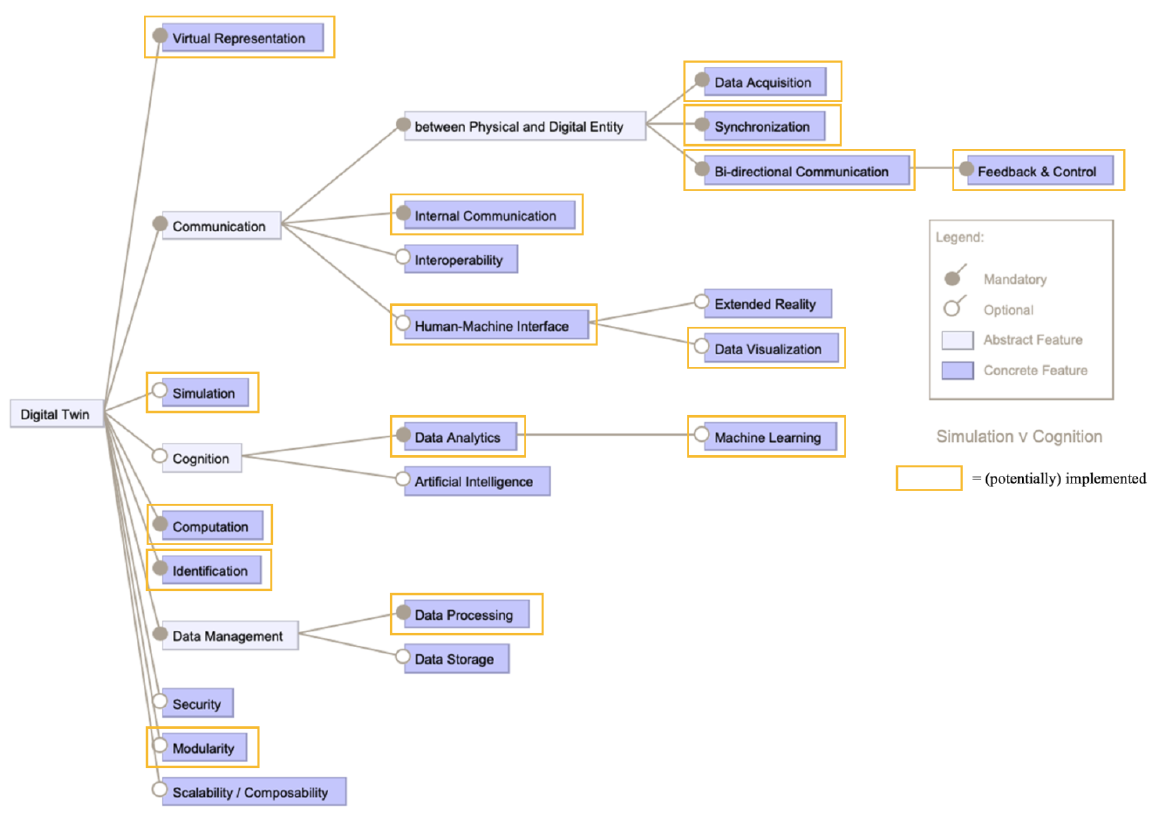}
    \caption{Feature Model of a DT for Temperature Monitoring in Welding Processes: 15 (Potentially) Implemented Features; Created with FeatureIDE (Eclipse).} 
    \label{fig:welding}
\end{figure}

In conclusion, effective monitoring may require the implementation of 15 DT features, as illustrated in Fig.~\ref{fig:welding}. 
This monitoring application facilitates bi-directional communication, enabling the use of a DT in place of a digital shadow.

\section{Discussion}
\label{sec:discussion}

This research sought to establish a GFM for DTs to address the absence of a systematic, cross-domain depiction 
of critical and supplementary characteristics in DT engineering. We, therefore, conducted a thorough literature analysis 
and mapping to identify fundamental features across several domains and integrated them into a structured model. 
In the following, we discuss the results and implications of our findings concerning our study questions and hypotheses, identify 
issues, and delineate prospective future endeavors.

\subsection{Breaking it Down: Answers to RQs}

In the following, we provide answers to our research questions from~\cref{sec:introduction}.

\subsubsection{RQ1: What are the DT problem spaces?}

DTs operate in complex environments with diverse challenges. The problem space refers to the key obstacles, 
constraints, and considerations when designing and implementing DTs. This includes issues such as data 
integration, real-time processing, security, interoperability, and life cycle management. Understanding 
the problem space is crucial to identifying the necessary features and ensuring the DT’s effectiveness in 
real-world applications. As part of this, we have identified twelve challenges (cf.~\cref{ssec:problem-space}) 
that are crucial, rooted in the DT problem spaces, which further guided the development of the GFM. 

\subsubsection{RQ2: What are the DT design spaces?}

The design space of DTs refers to the core concepts and components that define a DT’s structure, functionality, 
and interactions. These include monitoring, state representation, visualization, intention (goal-setting), 
situation detection, decision support, behavior modeling, simulation, adaptation, and control. Aligning these 
concepts with industry requirements ensures that DTs can effectively model and interact with their physical 
counterparts. Addressing these, we have identified ten core concepts rooted in the design spaces (cf.~\cref{ssec:design-space}) 
which are refined along concrete features in our GFM.

\subsubsection{RQ3: What are the mandatory features of DTs across domains?}

For the DM we discovered 5 mandatory features, for the DS 11 and for the DT GFM 13 respectively. 
Mandatory features are those essential for any DT, regardless of the domain. Our analysis of DT literature 
has confirmed that there are a number of domain-agnostic features for DTs, DSs and DMs. These typically include bi-directional 
communication, virtual representation, simulation, real-time data acquisition, and data integration. Without these 
core functionalities, a DT cannot effectively function as a real-time digital counterpart of a physical system. 
This study identified these fundamental features to establish a baseline for DT development.

\subsubsection{RQ4: What are the optional features of DTs across domains?}

For the DM we discovered 13 optional features, and for the  DS and DT GFM 11. Optional features vary based on the 
application domain and the intended use case of the DT. Examples include 
artificial intelligence (AI) integration, predictive analytics, human-machine interaction, modularity, and 
domain-specific customizations. These features enhance a DT’s capabilities but are not always necessary for 
basic functionality. Their inclusion depends on industry needs, cost constraints, and technological maturity.

\subsubsection{RQ5: What is the DT solution space?}

The solution space encompasses the possible implementations and feature configurations of DTs to address the 
challenges identified in the problem space. This includes different architectures, data processing techniques, 
levels of fidelity, and deployment strategies. By mapping problem spaces to design spaces, the study provides 
guidance on how to build effective DTs that meet the needs of various industries. The GFM is the blueprint 
for feature selection given the problem and design spaces.

\subsection{Validation of the Hypotheses}

To assess the validity of our proposed GFM for DTs, we conducted a structured validation process. The primary 
objective was to evaluate whether the GFM effectively captures the mandatory and optional features of DTs across 
different domains. This validation is based on an empirical approach, analyzing the model’s applicability in 
real-world DT implementations, viz. (cf.~\cref{sec:application}):
\begin{itemize}
    \item Emergency Management DT: Focused on real-time monitoring and response coordination
    \item Vehicular DT: Evaluated features such as simulation and data-driven control
    \item Manufacturing DT: Assessed life cycle management and predictive maintenance capabilities
\end{itemize}
As shown in~\cref{sec:application}, the cases demonstrate that the feature model provides a consistent 
framework adaptable to multiple DT implementations.

Based on this validation, the results support \textbf{H1} and refute \textbf{H0}. The GFM successfully generalizes 
mandatory and optional DT features across domains, reinforcing its relevance for DT design, implementation, and 
verification and validation.

\subsection{Implications~\&~Challenges}

The results of this study have substantial ramifications for the advancement and utilization of DTs in various 
domains. By establishing a GFM for DTs, our work introduces a structured framework that facilitates the design, 
development, and validation of DTs. Our GFM significantly contributes to the advancement of the standardization of DTs. 
The GFM establishes a systematic classification by differentiating between mandatory and optional features, hence 
enhancing comparability across DT implementations. This yields a standardized language for DT construction, promoting 
uniformity in execution and evaluation. Moreover, integrating the GFM with established DT maturity frameworks enables 
stakeholders to assess their present level of DT adoption and strategize future enhancements more efficiently.

A significant implication of our work is its capacity to enable model-driven development of DTs. The organized structure 
of the GFM establishes a basis for automated model-to-model and model-to-text transformations, enhancing the efficiency 
of the development process and diminishing the complexity of DT implementation. The GFM facilitates feature-based configuration 
through a clearly delineated set of features, allowing DTs to be customized for particular industry requirements while 
preserving a cohesive and scalable framework. This facilitates more effective decision-making across the DT development 
life cycle, enabling stakeholders to methodically identify important features for their use case and distinguish them 
from optional ones.

In addition to facilitating the creation, the GFM is essential for the verification and validation of DTs. By clearly delineating essential 
components, the GFM facilitates organized test case creation and compliance verification, guaranteeing that DT implementations 
adhere to established requirements. This method improves reliability and mitigates dangers linked to incomplete or 
inconsistent DT configurations. The systematic arrangement of features aids in ensuring that implemented functionality 
corresponds with planned behaviors, hence enhancing the robustness of the DT validation process.

Notwithstanding these benefits, certain problems must be confronted to guarantee the proper usage of the GFM. 
A primary challenge is reconciling generality with domain-specific needs. Although the model is intended for application 
across diverse domains, distinct sectors possess specific limits and requirements that may not consistently conform to 
a generalized framework. DTs in healthcare and critical infrastructure systems, for example, frequently necessitate enhanced 
security, regulatory compliance, and real-time limitations that go beyond such a generalized model.

A further problem pertains to the integration with current undocumented DT solutions. Numerous sectors have already 
developed DT frameworks, frequently customized to their own operational requirements. Aligning existing infrastructures to 
the GFM might necessitate a meticulous evaluation of compatibility with historical systems and data models. Organizations may encounter 
challenges in aligning their existing DTs with the GFM, especially if their systems were designed in an unstructured manner. 
Addressing these alignments might also necessitate modifications and domain-specific enhancements to the GFM.

\vspace{.25cm}
Our suggested feature model represents a substantial advancement in the standardization and structure of DT development. 
Although it presents distinct benefits regarding comparability, model-driven engineering, and validation, its extensive 
adoption will hinge on its ability to address domain-specific requirements and align with existing solutions. Confronting 
these issues will be essential to guarantee that the GFM functions as an effective and scalable instrument for directing 
DT development across various domains.
%%MV_stop
\subsection{Threats to Validity}

The suggested GFM for DTs offers a systematic framework for identifying mandatory and optional features in DTs; nevertheless, 
potential limitations and risks towards its validity must be acknowledged. These threats stem from methodological decisions, 
the study's scope, and the intrinsic complexity of DT engineering.

\emph{Construct validity:} A further critical factor is construct validity, which pertains to whether the specified features  
accurately reflect the characteristics of DT. Given that DTs are utilized throughout a wide array of businesses, the feature 
descriptions in the literature may represent domain-specific interpretations rather than universal principles. The differentiation 
between mandatory and optional features was established through cross-domain comparisons and conceptual congruence with current 
DT maturity models; nevertheless, this classification is still open to interpretation. Features classified as optional may be 
essential in specific applications, while those considered mandatory may not always be required in simpler DTs.

\emph{Internal Validity:} A major challenge to internal validity is the dependence on literature-derived feature extraction. 
The SLR established the basis of the GFM, indicating that any biases in study selection or deficiencies in existing research 
may have affected the comprehensiveness of the detected features. Despite adhering to a stringent and well-defined process, with established 
inclusion and exclusion criteria, there remains a possibility that pertinent publications were missed, especially those not 
indexed in the chosen databases or written in languages other than English. Moreover, variations in language among studies may have 
resulted in certain aspects being classified differently or omitted due to inconsistent definitions.

\emph{External Validity:} External validity poses an additional hurdle, especially regarding the generalizability of the 
proposed feature model. Although attempts were made to ensure broad application across various domains, the cases 
employed for validation concentrated on certain industrial examples, including emergency management, vehicular systems, 
and manufacturing. These examples offer robust empirical evidence; nonetheless, they do not cover the complete spectrum of 
DT applications, including those in healthcare, energy management, or urban infrastructure. Consequently, although the GFM 
encompasses a broad spectrum of properties, it may necessitate enhancements or expansions when utilized in sectors not 
explicitly addressed in this work. Subsequent assessments incorporating a broader array of actual DT implementations may 
enhance the model's generalizability.

\vspace{.25cm}
A final threat to validity pertains to the concurrent evolution of DT technology. DTs are still evolving, incorporating 
novel functionalities such as artificial intelligence-driven automation, edge computing, and blockchain-based security. As 
technology advances, several features highlighted in this study may become outdated, while new critical features may arise. 
The GFM provides a current overview of DTs, although its enduring significance will rely on regular updates and enhancements 
to remain consistent with technical progress and evolving best practices.

Despite these threats, the systematic methodology, interdisciplinary approach, and validation via cases establish a robust 
foundation for the GFM. Recognizing the above threats to validity enables a more prudent interpretation of our results, 
while also emphasizing opportunities for future research and enhancement. Addressing these issues via ongoing validation, 
extensive case analyses, and iterative enhancements will be essential to guarantee that the GFM remains a dependable 
and flexible instrument for DT engineering.

\section{Conclusion}
\label{sec:conclusion}

In this article, we have developed and validated a GFM for DTs, providing a systematic framework for understanding the 
problem spaces, design spaces, and solution space in DT engineering. Through a comprehensive SLR, 
we identified, extracted, and categorized both mandatory and optional features that form the foundation of DTs across diverse 
application areas. Our findings reveal that while specific implementation details may vary, the core set of features 
remains largely consistent across domains, underscoring the universal applicability of the proposed model.

Moreover, the GFM integrates insights from existing maturity models, design frameworks, and domain-specific applications, 
thereby establishing a cohesive, feature-based approach to DT engineering. This framework enhances the structured 
development, evaluation, and validation of DTs, facilitating their scalability and adaptability across multiple 
domain contexts, including manufacturing, healthcare, smart cities, and aerospace. By explicitly distinguishing between 
mandatory and optional features, the model provides clear guidelines for selecting appropriate capabilities 
based on specific application needs, improving both the precision and adaptability of DT implementations.

Despite the robust methodological approach and empirical validation across diverse use cases, certain challenges 
and limitations must be acknowledged. Construct validity concerns arise due to the inherent variability in feature 
descriptions across domains, which may reflect domain-specific nuances rather than universal characteristics. Internal 
validity is influenced by the reliance on literature-derived feature extraction, which may introduce selection biases 
or overlook emerging features not yet widely documented. Furthermore, the external validity of the GFM is limited by 
the scope of validation use cases, which, although comprehensive, do not fully capture the breadth of potential DT 
applications in areas such as energy management and urban infrastructure.

An additional consideration is the rapid evolution of DT technology. Emerging advancements such as artificial 
intelligence-driven automation, edge computing, and blockchain integration will likely necessitate ongoing 
revisions to the feature model. Future work should prioritize continuous model updates, incorporating new 
technologies, and addressing evolving domain requirements. Moreover, expanding the model's validation through 
broader case studies and real-world implementations will further enhance its generalizability and practical 
utility.

To ensure the model's sustained relevance and applicability, future research should also explore means for integrating automated feature selection processes using AI-driven methodologies. Such approaches could help to dynamically adapt DT 
configurations in response to real-time performance data, improving both efficiency and responsiveness. Furthermore, 
the development of engineering tools that embed the GFM into DT development workflows will be crucial for 
bridging the gap between theoretical models and practical deployment.

In conclusion, the proposed GFM for DTs represents a significant contribution to the systematic understanding and 
implementation of DTs. By offering a comprehensive, adaptable, and validated framework, it provides researchers and 
practitioners with a robust foundation for designing, evaluating, and refining DTs across a range of application 
domains. Addressing the identified challenges through continuous validation and technological integration will be 
essential for ensuring the GFM remains a valuable tool.

\bibliographystyle{elsarticle-num}
\bibliography{references}

\end{document}